\documentclass[showpacs,twocolumn,pra,superscriptaddress,nofootinbib]{revtex4-1}
\pdfoutput=1

\usepackage[dvipsnames]{xcolor}
\usepackage{amsfonts}
\usepackage{amsbsy}
\usepackage{bm}
\usepackage{color}
\usepackage{array}
\usepackage{dcolumn}
\usepackage{tensor}
\usepackage{braket}
\usepackage{eufrak}
\usepackage{mathrsfs}
\usepackage{slashed}
\usepackage{amsmath,amssymb,graphicx}
\usepackage{mathrsfs}
\usepackage{bbm}
\usepackage{bbold}
\usepackage{hyperref}
\usepackage{slashed}
\usepackage[normalem]{ulem} 
\usepackage{tikz}
\usepackage{verbatim}
\usepackage[utf8]{inputenc}

\DeclareFontFamily{U}{mathb}{}
\DeclareFontShape{U}{mathb}{m}{n}{
  <-5.5> mathb5
  <5.5-6.5> mathb6
  <6.5-7.5> mathb7
  <7.5-8.5> mathb8
  <8.5-9.5> mathb9
  <9.5-11.5> mathb10
  <11.5-> mathbb12
}{}

\newcommand{\be}{\begin{equation}}
\newcommand{\ee}{\end{equation}}
\newcommand{\bi}{\begin{itemize}}
	\newcommand{\ei}{\end{itemize}}
\newcommand{\bea}{\begin{eqnarray}}
\newcommand{\eea}{\end{eqnarray}}
\newcommand{\tr}{\text{tr}\,}
\newcommand{\trm}[1]{\textrm{#1}}

\newcommand{\eqnref}[1]{Eq. (\ref{#1})}
\newcommand{\figref}[1]{Fig. \ref{#1}}

\newcommand{\ba}{\begin{align}}
\newcommand{\ea}{\end{align}}

\newcommand{\eps}{\varepsilon}

\newcommand{\nn}{\nonumber}

\newcommand{\sfrac}[2]{{\textstyle \frac{#1}{#2}}}

\newcommand{\ud}{\mathrm{d}}
\newcommand{\LCm}{{\scriptscriptstyle -}} 
\newcommand{\LCp}{{\scriptscriptstyle +}}
\newcommand{\LCpm}{{\scriptscriptstyle \pm}}

\newcommand{\LCperp}{{\scriptscriptstyle \perp}}

\newcommand{\mcI}{\mathcal{I}}

\newcommand{\mcL}{\mathcal{L}}
\newcommand{\mcM}{\mathcal{M}}

\newcommand{\mcO}{\mathcal{O}}

\newcommand{\mcR}{\mathcal{R}}
\newcommand{\mcS}{\mathcal{S}}
\newcommand{\mcT}{\mathcal{T}}

\newcommand{\bbP}{\mathbb{P}}

\newcommand{\lp}{{\ell^\prime}}
\newcommand{\pp}{{p^\prime}}
\newcommand{\qp}{{q^\prime}}

\newcommand{\vphi}{\varphi}
\newcommand{\half}{\frac{1}{2}}

\begin{document}


\title{
    The locally monochromatic approximation to QED in intense laser fields
}

\author{T. Heinzl}
\email{tom.heinzl@plymouth.ac.uk}
\affiliation{Centre for Mathematical Sciences, University of Plymouth, Plymouth, PL4 8AA, United
Kingdom}

\author{B. King}
\email{b.king@plymouth.ac.uk}
\affiliation{Centre for Mathematical Sciences, University of Plymouth, Plymouth, PL4 8AA, United
Kingdom}

\author{A. J.~MacLeod}
\email{alexander.macleod@plymouth.ac.uk}
\affiliation{Centre for Mathematical Sciences, University of Plymouth, Plymouth, PL4 8AA, United
Kingdom}

\date{\today}
\begin{abstract}
We derive an approximation to QED effects in intense laser fields which can be employed in laser-particle collisions.
Treating the laser as a plane wave of arbitrary intensity, we split the wave
into fast (carrier) and slow (envelope) modes. We solve the interaction dynamics
exactly for the former while performing a local expansion in the latter. This
yields a `locally monochromatic' approximation (LMA), which we apply to
nonlinear Compton scattering in circularly- and linearly-polarised backgrounds
and to nonlinear Breit-Wheeler pair production. We provide the
explicit link between the LMA and QED, and benchmark against exact QED results.
The LMA is particularly useful for high-energy, intermediate-intensity collisions, where, unlike the `locally-constant field' approximation, the LMA correctly describes the position and amplitude of harmonic features and exactly reproduces the low energy limit. We show that in the limit of high-intensity and large harmonic order, the locally-constant field approximation is recovered from the LMA.
\end{abstract}

\maketitle

\section{Introduction \label{sec:Intro}}
%

There is a growing interest in experimentally verifying the predictions of
quantum electrodynamics (QED) in the strong field, high-intensity, regime. To
access this regime in experiment, two requirements must be met: (i) an
electromagnetic field is present which is sufficiently intense so that many
field quanta participate in a given process; (ii) the momentum transfer (recoil)
in scattering is large enough that the quantum nature of processes is manifest.
Upcoming laser facilities such as ELI-Beamlines~\cite{eli-beams},
ELI-NP~\cite{eli-np}, and SEL~(see~\cite{Danson:2019} for an overview) will
reach field strengths to fulfil requirement~(i). One way to fulfil~(ii) is to
use laser wakefield accelerated particles, recent successes of which include the
generation of positron beams in the lab~\cite{sarri15} and measurement of
quantum signals of radiation reaction~\cite{Cole:2017zca,Poder:2018ifi}. 

Background electromagnetic field strength can be quantified using an intensity
parameter, $\xi$, equivalent to the work done by the background over a Compton
wavelength, in units of the background photon energy. When $\xi\sim O(1)$, the
standard approach of treating the background in perturbation theory fails,
because this assumes that processes are more probable when \textit{fewer}
background photons are involved. When $\xi\gg1$, an alternative approximation is
often employed, in which the instantaneous rate for processes in a constant
(`crossed') plane wave background (treated without recourse to perturbation
theory) is integrated over the classical trajectories of the scattered
particles. This ``locally-constant field approximation'' (LCFA)
\cite{Nikishov:1964zza,Ritus:1985,king15d,DiPiazza:2017raw} has the particular
advantage that it can be applied to arbitrary external fields. Therefore, when
used in conjunction with a classical Maxwell field equation solver, it can be
employed in situations for inhomogeneous backgrounds. The locally-constant field
approximation is almost exclusively the method by which QED processes in intense
fields are added to laser-plasma simulation codes
\cite{nerush11,elkina11,ridgers12,king13a,bulanov13,ridgers14,Gonoskov:2014mda,blackburn14,gelfer15,jirka16,gonoskov17,efimenko19}.
It has recently been extended in several respects, by including higher
derivative corrections \cite{Ilderton:2018nws,DiPiazza:2018bfu,King:2019igt},
analysing simple, non-constant, fields in Schwinger pair
production~\cite{aleksandrov19} and extending it to previously neglected
processes~\cite{Ilderton:2019bop,Tang:2019ffe}. 

An alternative approach to probe the strong-field regime of QED is to use a
conventional particle accelerator to fulfil the energy condition (ii), and a
less intense laser to fulfil the field condition (i). This was demonstrated by
the landmark E144 experiment~\cite{Bamber:1999zt} which investigated photon
emission~\cite{Bula:1996st} and pair production~\cite{Breit:1934zz,Burke:1997ew}
in the weakly nonlinear regime.  Using modern high-intensity laser systems, this
form of experiment will be performed at E320 at FACET-II and at LUXE
\cite{Abramowicz:2019gvx} at DESY, to measure QED in the highly nonlinear,
non-perturbative regime, which was out of reach for E144.  These experiments
will access the intermediate intensity regime $\xi\sim O(1)$, where the
locally-constant field approximation breaks down and fails to capture
experimental observables such as the harmonic structure in
spectra~\cite{Chen:1998,sakai15,Khrennikov15}.

To address this problem we derive here, from QED, the ``locally monochromatic
approximation'' (LMA). Because the LMA is based upon a perturbation around a monochromatic background, it is not suitable for intense laser-matter collisions where a plasma is generated. Instead, it \emph{complements} the locally-constant-field approximation by covering the regime of high-energy and intermediate intensity where the LCFA becomes invalid. As usual, we assume that the laser background is well defined and backreaction~[\cite{seipt17,Ekman:2020vsc} can be neglected to a first approximation.
Rather than taking the constant crossed field result to be fundamental and the basis of the approximation, the LMA builds upon the monochromatic result, which is more specific to propagating fields such as laser pulses. One can show that both field configurations are 'null'  (characterised by vanishing field invariants) and thus have the \emph{same} degree of symmetry so that the dynamics becomes maximally super-integrable in either case \cite{Heinzl:2017zsr,Heinzl:2017blq}.

Various numerical codes have already been implemented that include the QED effects of nonlinear Compton scattering and nonlinear Breit-Wheeler pair-creation, by using an ``instaneously monochromatic'' rate that samples a non plane-wave field around the probe particles. Examples include the simulation code to support the SLAC E144 experiment \cite{Bamber:1999zt}, CAIN \cite{cain1} and IP Strong \cite{hartin18}, which has lately been used to provide simulation support for the planning of the LUXE experiment \cite{Abramowicz:2019gvx}.

In this paper, we formalise the LMA and identify the approximations necessary to derive it from QED. We find the LMA treats the fast dynamics related to the carrier
frequency of the plane wave exactly, but uses a local expansion to describe the
slow dynamics associated with the pulse envelope.  This combines the
slowly-varying envelope
approximation~\cite{Narozhnyi:1996qf,McDonald:1997,Seipt:2010ya,Seipt:2014yga,Seipt:2016rtk}
with the locally-constant field approximation, improving upon both. It
captures features to which the locally-constant field approximation is blind,
yet because it is still an explicitly local approximation, it can be added to
single-particle simulation codes. Furthermore, by benchmarking the LMA against
exact calculations in pulses, an additional feature in the mid-IR region of nonlinear Compton scattering will become apparent, which may provide an additional signal to be searched for in experiment.

The paper is organised as follows. In Sec.~\ref{sec:LMA} we outline the key
steps in deriving the LMA for a general first-order strong field QED process.
In Sec.~\ref{sec:FullQED} we give an outline of the numerical methods that form
the basis of our benchmarking against finite-pulse results. The LMA for
nonlinear Compton scattering is then compared to QED in circularly and linearly
polarised pulse backgrounds in Sec.~\ref{sec:NLC}. We demonstrate the validity
of the LMA for nonlinear Breit-Wheeler pair production in Sec.~\ref{sec:BW}. We
conclude in Sec.~\ref{sec:Summary}. In Appendix~\ref{app:NLCLMA}, a detailed
derivation of the LMA for nonlinear Compton scattering in a circularly polarised
background is presented and in  Appendix~\ref{app:Smalls} we include an
alternative derivation of the infra-red\footnote{Here and throughout, we use
`infra-red' to denote low \textit{lightfront} energy $n\cdot P$, for $n$ the
laser propagation direction and $P$ any given particle momentum. This is a
natural variable in plane wave calculations.} limit of nonlinear Compton
scattering, demonstrating also that the correct limit is trivially reproduced
from the LMA. Finally, in Appendix~\ref{app:LCFA}, we show that the
locally-constant field approximation can be recovered as a high-intensity limit
of the LMA.

%
\section{Outline of the locally monochromatic approximation\label{sec:LMA}}
%

Let the gauge potential of the background, $a_\mu(\varphi)$, depend only on the
phase $\varphi = k \cdot x$, with $k$ being the wave four-vector. We will work
in lightfront coordinates $x = (x^\LCp,x^\LCm,\bm{x}^\LCperp)$ where $x^{\LCpm}
= x^0 \pm x^3$ and $\bm{x}^\LCperp = (x^1, x^2)$. Here $x^\LCp$ is lightfront
time while $x^{\LCm}$ and $\bm{x}^{\LCperp}$ are called the longitudinal and
perpendicular directions, respectively~\cite{Heinzl:2000ht}. With this notation,
the wave vector of the background $k_\mu = \delta_\mu^+ k_+$, and $\varphi = k_+
x^+$.  The scattering amplitude, $S_{fi}$, for an incoming electron with
on-shell momentum $p$, $p^{2}=m^{2}$, is then calculated using the Volkov
wavefunction~\cite{Volkov:1935zz},
\begin{align}\label{def:Volkov}
\Psi_p(x)
=
\left(
1
+
\frac{\slashed{k} \slashed{a}(\varphi)}{2 k \cdot p}
\right)
u_p
e^{- i S_p(x)} \; .
\;
\end{align}
In the exponent,  $S_p(x)$ is the classical action for an electron in a plane
wave background,
\begin{align}\label{def:ClassAct}
S_p(x)
= 
p \cdot x
+
\int^{\varphi}_{- \infty}
\frac{2 p \cdot a(t) - a^2(t)}{2 k \cdot p}dt
.
\end{align}

The scattering amplitude $S_{fi}$ in a plane wave background can then be written
as
\begin{align}\label{def:Amplitude}
S_{fi}
=&
(2\pi)^3
\delta^3_{-,\perp}(p_{\text{in}} - p_{\text{out}})
\mcM,
\;
\end{align}
with an invariant amplitude $\mcM$. Due to the non-trivial structure of the
background, overall momentum conservation (encoded in the delta functions) only
holds in three directions, $\{-,\perp\}$.

A closed form solution for phase integrals such as (\ref{def:ClassAct}) is only
known for some special cases of the background field, for example infinite
``monochromatic'' plane waves (see e.g.~\cite{Ritus:1985} for extensive
applications).  Beyond these solutions, one can turn to a numerical approach or
employ an approximation. The slowly varying envelope approximation is known to
simplify the classical action (\ref{def:ClassAct}) occurring in the exponent and
hence make the phase integrations
tractable~\cite{Narozhnyi:1996qf,McDonald:1997,Seipt:2010ya,Seipt:2014yga,Seipt:2016rtk}. 
It is applied as follows. Let the pulse $a_\mu(\varphi)$ have the form
\begin{align}\label{def:Gauge}
a^\mu(\varphi)
=&
m \, \xi \, 
f\Big(\frac{\varphi}{\Phi}\Big)
\big(
\varepsilon^\mu
\cos\delta
\cos\varphi
+
\bar{\varepsilon}^\mu
\sin\delta
\sin\varphi
\big)
\;,
\end{align}
where $\xi$ is the dimensionless Lorentz and gauge invariant measure of the
field intensity~\cite{Heinzl:2008rh}, $f(\varphi/\Phi)$ is the pulse envelope
with phase duration $\Phi$ and $\varepsilon^\mu$, and $\bar{\varepsilon}^\mu$
are polarisation directions satisfying $\varepsilon^2 = \bar{\varepsilon}^2 =
-1$ and $\varepsilon \cdot \bar{\varepsilon} = k \cdot \varepsilon = k \cdot
\bar{\varepsilon} = 0$.  The parameter $\delta \in (0,\pi/2)$ determines the
polarisation of the pulse; $\delta = 0$  for linear polarisation along
$\varepsilon$, $\delta = \pi/2$  for linear polarisation along
$\bar{\varepsilon}$ and $\delta = \pi/4$  for circular polarisation\footnote{We
make implicit a normalisation factor in the gauge potential (\ref{def:Gauge})
such that $\text{Max}[a_\mu(\varphi)/(m \xi)] = 1$.}.  We consider the pulse
envelope to be asymptotically switched on and off,
$\lim_{\varphi\to\pm\infty}f(\varphi) = 0$.

The slowly varying envelope approximation assumes that the pulse duration $\Phi$
is sufficiently long that terms of order $\mcO(\Phi^{-1})$ can be neglected.
(Higher orders can in principle be included in the approximation but they will
lead to a more complicated result that takes longer to numerically evaluate and,
as we shall see, the leading order terms will already be sufficient to reproduce
the main features of spectra.) As a result, derivatives of the envelope with
respect to the phase can be neglected, because they are of the form
$df(\varphi/\Phi)/d\varphi \sim \Phi^{-1}f'(\varphi/\Phi)$. In other words, the
envelope varies slowly compared to the fast dynamics of the carrier frequency.
The practical benefit of this is that we can simplify the classical action
(\ref{def:ClassAct}). More explicitly, the classical action will have terms both
linear and quadratic in the field envelope. In all terms involving both fast and
slow oscillations, we integrate by parts, picking up terms of order
$\mcO(\Phi^{-1})$ which we neglect, and so remove the integrals from
(\ref{def:ClassAct}). This gives us, for the possible linear terms arising,
\begin{align}\label{eqn:LinearIntApprox}
\int^{\varphi}_{-\infty}
\!
\ud \psi
\;
f\Big(
\frac{\psi}{\Phi}
\Big)
\big\{
\cos \psi
,
\sin \psi
\big\}
\simeq
f\Big(
\frac{\varphi}{\Phi}
\Big)
\big\{
\sin\varphi
,
-
\cos\varphi
\big\}
\,,
\end{align}
and for the possible quadratic terms
\begin{align}\label{eqn:QuadIntApprox}
&
\int^{\varphi}_{-\infty}
\!
\ud \psi
\;
f^2\Big(
\frac{\psi}{\Phi}
\Big)
\big\{\cos^2 \psi
,
\sin^2 \psi
\big\}
\nonumber\\
&\simeq
\frac12
f^2\Big(
\frac{\varphi}{\Phi}
\Big)
\Big\{
\big(
\varphi
+
\sin\varphi
\cos\varphi
\big)
,
\big(
\varphi
-
\sin\varphi
\cos\varphi
\big)
\Big\}
\,.
\end{align}
For the particular case of a circularly polarised background, there arises a
term containing only slow oscillations (the integral of $f^2$ without
trigonometric functions), which must be approximated by different means (see
below). With these approximations, the background-dependent parts of the
classical action can always be put in the form
\begin{align}\label{eqn:ClassActApprox}
S_p(x)
&\simeq
G\left(\vphi,\frac{\varphi}{\Phi}\right)
+
\half 
\alpha\Big(\frac{\varphi}{\Phi}\Big)
\big[u(\varphi) - u^{-1}(\varphi)\big]
\nonumber\\
&+
\half 
\beta\Big(\frac{\varphi}{\Phi}\Big)
\big[v(\varphi) - v^{-1}(\varphi)\big] \; .
\end{align}
The functions $\alpha$ and $\beta$ are purely slowly-varying functions of the
phase $\varphi$.  The functions $u(\varphi)$ and $v(\varphi)$ are of the form
$\exp(ic\vphi)$, for $c\in\{1,2\}$. Note the similarity of the form of the
exponent with the generating function for the Bessel function of the first kind,
\begin{equation}\label{def:Generating}
\exp \left\{
\sfrac{1}{2}
z\left( \frac{\varphi}{\Phi} \right)
\left[u(\varphi) - u^{-1}(\varphi)\right] \right\}
=\!
\sum_{n \in \mathbb{Z}}
u^{n}(\varphi)
J_{n}\!\left[z\left(\frac{\varphi}{\Phi}\right)\right]\!.
\end{equation}
This was recognised and exploited in~\cite{Narozhnyi:1996qf} and essentially
gives a generalisation of the infinite monochromatic field
results~\cite{Berestetsky:1982aq,Ritus:1985} to the case where the argument of
the Bessel function now depends slowly on the phase.  There will also appear
rapidly oscillating terms in the pre-exponent, but these can be incorporated by
differentiating (\ref{def:Generating}) with respect to $z$ and combining terms.
The scattering amplitude will thus be defined in terms of harmonics, represented
by the sum over integers~$n$ in (\ref{def:Generating}). 

So far everything has been typical for the application of the slowly-varying
envelope approximation in the strong-field QED
literature~\cite{Narozhnyi:1996qf,McDonald:1997,Seipt:2010ya,Seipt:2014yga,Seipt:2016rtk}.
It is at this point that we take the further step of performing a local
expansion in the phase variables to arrive at a local ``rate'' which can be
implemented in one-particle numerical simulations.

To define the local expansion, we will concentrate on single (dressed) vertex
``one-to-two'' processes:  nonlinear Compton scattering and nonlinear
Breit-Wheeler pair production. The amount of literature on these processes has
become too large to be cited here in full; regarding nonlinear Compton
scattering see \cite{Nikishov:1964zza,Brown:1964zzb,Goldman:1964} for the
original papers, \cite{Ritus:1985,Ehlotzky:2009} for reviews and
\cite{Harvey:2009ry,Boca:2009zz,Mackenroth:2010jr,Heinzl:2009nd,Seipt:2010ya}
for a selection of more recent results. Nonlinear Breit-Wheeler pair creation
was first discussed in \cite{Toll:1952rq,Reiss:1962,Nikishov:1964zza}, while the
study of finite size effects was initiated in \cite{Heinzl:2010vg}. Both
processes were observed (at mildly nonlinear intensities) by the SLAC E144
experiment \cite{Bula:1996st,Burke:1997ew,Bamber:1999zt}. For the two examples
to be considered, the reduced amplitude $\mcM$ in (\ref{def:Amplitude}) will
have one phase integral, and after applying the slowly-varying envelope
approximation, will be defined in  terms of an infinite sum over the harmonic
order $n$, i.e.
\begin{align}
\mcM
=
\sum_{n = - \infty}^{\infty}
\int
\!
\ud \varphi
\;
\mcM_{n}(\varphi)
.
\end{align}
Squaring the amplitude for the probability, we will have something of the form,
\begin{align}
\bbP
\sim
\sum_{n,n^\prime = - \infty}^{\infty}
\int
\ud \Omega_{\text{LIPS}}
\int
\!
\ud \varphi
\,
\ud \varphi^\prime
\;
\mcM_{n}^{\dagger}(\varphi)
\mcM_{n^\prime}(\varphi^\prime)
\;,
\end{align}
i.e., a double infinite sum over harmonic orders, two phase integrals, and an
integration over the Lorentz invariant phase space of the process,
$\ud\Omega_{\text{LIPS}}$. 

Now we perform a local expansion of the probability, in analogy to the
locally-constant field approximation (see
e.g.~\cite{Nikishov:1964zza,Ritus:1985,king15d,DiPiazza:2017raw}). We make a
change of variables to the sum and difference of phases,
\begin{align}\label{def:Phase}
\phi
=
\frac{1}{2}
\big(
\varphi + \varphi^\prime
\big)
\;,
\quad
\theta
=
\varphi - \varphi^\prime
\;.
\end{align}
Terms in the probability are then expanded in a Taylor series in $\theta \ll 1$,
and the slowly-varying envelope approximation is then applied to all derivatives
of the pulse envelope, giving
\begin{align}
f\Big(\frac{\varphi}{\Phi}\Big)
\approx
f\Big(\frac{\varphi^\prime}{\Phi}\Big)
\approx
f\Big(\frac{\phi}{\Phi}\Big)
\;.
\end{align} 
This allows the $\ud \theta$ integrals to be performed, and the probability
takes the form
\begin{align}
\bbP
=
\int
\!\ud\phi
\;
\mcR(\phi)
,
\end{align}
where $\mcR(\phi)$ is interpreted as a local ``rate''\footnote{
    In general $\mcR(\phi)$ will contain infinite sums over harmonic orders,
    and a number of final state momentum integrals. The aim is to do as many
    of these final state momentum integrals as possible.  Despite the added
    complexity which arises from retaining a slowly varying dependence on
    the phase variable $\phi$, the number of final state integrals that can
    be performed is the same in the LMA as for a first order process
    in an infinite monochromatic plane wave
    \cite{Berestetsky:1982aq,Ritus:1985} (see appendix~\ref{app:NLCLMA}).
}. 
For the processes of nonlinear Compton scattering and nonlinear Breit-Wheeler
pair production we can write:
\bea
\bbP_{\scriptstyle \textsf{LMA}}  \approx \int \ud\phi \, \mcR_{\scriptstyle{\textsf{mono}}}[\xi f(\phi/\Phi)]
\;, \label{eqn:exp1}
\eea
where $\mcR_{\scriptstyle{\textsf{mono}}}$ is the probability per unit phase of
the process in a monochromatic (infinitely long) plane wave. For a circularly
polarised background the LMA is exactly equal to the integral on the right-hand
side of (\ref{eqn:exp1}). For a linearly polarised background, it is not so
straightforward, as interference between different harmonic orders is included,
but we will find that, to a good approximation, both sides of (\ref{eqn:exp1})
are equal.

To conclude this outline of the LMA, we reiterate that the LMA is simply the
application of two well-known approximations in the strong-field QED literature,
the slowly varying envelope approximation and the ``local'' expansion in the
relative phase variable, $\theta$, carried out at the level of the probability.
For each term in the local expansion we apply the slowly-varying envelope
approximation, which reduces the complexity of the rates and allows us to
progress further analytically. What this means is that no further restrictions
have to be imposed on the pulse envelope beyond those required for the
slowly-varying envelope approximation to be valid, i.e. that the phase duration
$\Phi$ be sufficiently large that derivatives of the envelope can be safely
neglected.
Although the approximation has been used before for a circularly polarised
background~\cite{Titov:2019kdk}, as far as we are aware, this is the first
explicit derivation and benchmarking with the direct calculation from QED for a
plane-wave pulse. The monochromatic result is obtained from the LMA by taking
the infinite pulse limit $\Phi \to \infty$, i.e.  $f\to 1$. 
%
%
%
\section{Direct calculation from QED for a pulsed background \label{sec:FullQED}}
%

We wish to benchmark the LMA against the numerical evaluation of exact
expressions from high-intensity QED. We provide here the details of the
integration scheme used. For both nonlinear Compton scattering and nonlinear
Breit-Wheeler pair production in a plane-wave pulse, one can write the total
probability in the form $\bbP = \alpha \mathcal{I}/\eta$, where $\alpha$ denotes
the fine structure constant, $\eta=k\cdot P /m^2$ is the energy parameter of the
incoming particle (where $k$ is the light-like wave vector of the plane wave
background, $P$ is the four-momentum of the incoming particle) and $\mathcal{I}$
is a triple integral. $\mathcal{I}$ involves two phase integrals, $\phi$, $\theta$, and an integral over $s$, the fraction of the
incoming particle's light-front momentum, $P^{-}$, carried away by the emitted
particle. For nonlinear Compton scattering, this is of the form:
\bea
\mathcal{I} &=& \int_{-\infty}^{\infty}  \ud \phi \int_{0}^{1} \ud s  \left\{- \frac{\pi}{2}  \right. \nn \\
&&  \left. + \int_{0}^{\infty} \frac{\ud \theta}{\theta}\left[1 + h(a,s)\right]\sin\left[g(s)\theta\mu(\phi,\theta)\right] \right\} \, . \label{eqn:pulse1}
\eea
For the numerical calculation of the exact QED result, we are using the ``$i\epsilon$'' regularisation at the level of the probability (see e.g. \cite{dinu12}), as evidenced by the $\pi/2$ counter-term in \eqnref{eqn:pulse1}. The dependence on the field $a$ defined in
(\ref{def:Gauge}) resides in both $h(a,s)$ and in the Kibble mass
\cite{Kibble:1965zz,Kibble:1975vz} normalised by the electron mass:
\bea \label{eqn:Kibble}
\mu(\phi,\theta) = 1- \frac{1}{\theta}\int_{\phi - \theta/2}^{\phi + \theta/2} \frac{a^{2}}{m^{2}} + \bigg(
\frac{1}{\theta}
\int_{\phi - \theta/2}^{\phi + \theta/2} \frac{a}{m} \bigg)^2.\nn \\
\eea
In what follows we will outline some manipulations allowing for a
straightforward numerical integration of $\mathcal{I}$.  

\begin{figure}[!!t]
\centering 
\includegraphics[width=0.7\linewidth]{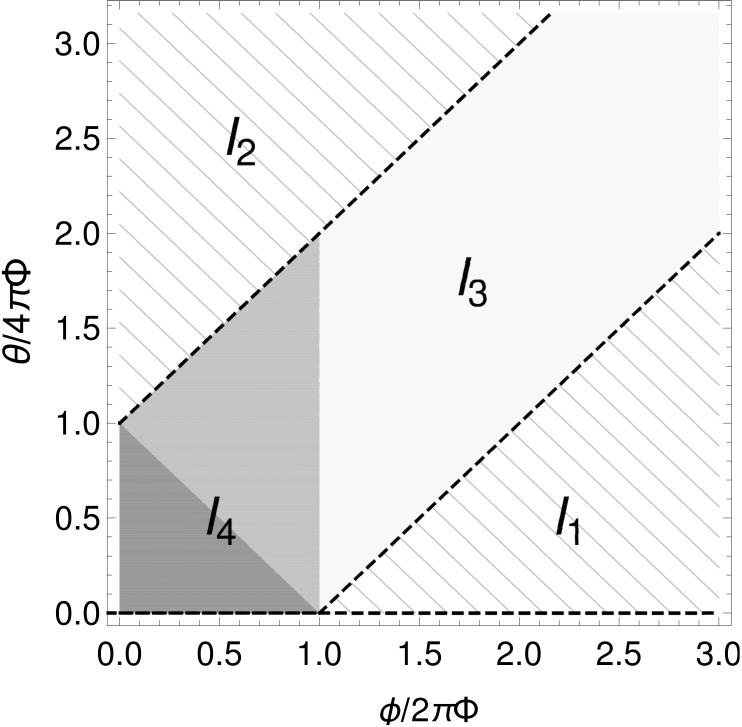}
\caption{Overview of the regions integrated over in the $\phi$-$\theta$ plane.
The non-striped regions are inside of the pulse: \mbox{$|\vphi'|<2\pi\Phi$}. The
dark subregion in the area covered by $I_{4}$ signifies $|\vphi|<2\pi\Phi$.} \label{fig:regionPlot1}
\end{figure}
The phase integration plane $(\phi,\theta)$ can be split naturally into
subregions where the integrand in (\ref{eqn:pulse1}) takes a specific form
according to the following two observations: First, the field-dependent function
$h(a,s)$ only has support for $a \ne 0$. Second, the Kibble mass becomes
phase-independent when $\phi \pm \theta/2$ obey certain inequalities (see
below). 

Suppose we consider a pulse envelope, $f(\vphi/\Phi)$, which is symmetric about
the origin with support $|\vphi|<L/2$. The example pulse shape we consider in
this paper is $f = \cos^{2}$, where the phase duration is  $L=\pi\Phi$ and the
pulse length parameter, $\Phi$, can be related to the number of cycles, $N$, via
$\Phi = 2N$. Using the symmetry of the integrand, we only have to consider the
first quadrant in the $(\phi,\theta)$-plane, which splits into the sub-regions
shown in \figref{fig:regionPlot1} such that $\mathcal{I} = \int ds \sum_{k=1}^4
I_k$.

To deal with the infinite numerical integation of a nonlinearly oscillating pure
phase term, we first rewrite the regularisation factor as
\[
\frac{\pi}{2} = \int_{0}^{\infty} \frac{\ud\theta}{\theta}\,\sin K\theta,
\]
which is independent of the choice of the constant factor $K$. In order to make
for a simpler numerical evaluation, we choose $K=g(s)$, allowing us to combine
it with the other infinite phase term in (\ref{eqn:pulse1}). (Other choices are
useful in other circumstances, see for example~\cite{Dinu:2013hsd} and
(\ref{eqn:NLCdif}) in the appendix.) Using this trick, we find that the first
integral vanishes,
\bea
I_1 = \int_{2\pi\Phi}^{\infty} \ud\sigma \int_{0}^{2(\phi-2\pi\Phi)} \frac{d\theta}{\theta}  & & \left\{ -\sin\left[g(s)\theta\right] \right. \nn  \\ && \left. + \sin\left[g(s)\theta\mu(\phi,\theta)\right]\right\} = 0.\nn
\eea
This can be shown by noting that $\lim_{a\to 0}\mu(\phi,\theta) = 1$, and in
this phase region the pulse has no support. This is because terms depending on
the potential, $a(\vphi)$, $a(\vphi')$ are zero unless:
\[|\vphi|=|\phi+\theta/2|<2\pi\Phi\quad\trm{or}\quad|\vphi'|=|\phi-\theta/2|<2\pi\Phi.\]
In contrast, the integral $I_{2}$ over the region where the pulse is yet to pass
through, is non-zero:
\bea
I_2=  \int_{0}^{\infty} \ud\phi \int_{2(2\pi\Phi+\phi)}^{\infty} \frac{\ud\theta}{\theta}& & \left\{ -\sin\left[g(s)\theta\right] \right. \nn  \\ && \left. + \sin\left[g(s)\theta\mu(\phi,\theta)\right]\right\}  \ne 0 . \nn
\eea
Nevertheless, it may be calculated analytically by noting that the combination
$\theta \mu(\phi,\theta)$ accumulates a constant total phase, $\theta \mu \to
\theta + \theta_{\infty}$, when the probe particle traverses the pulse and
continues to propagate in vacuum. Explicitly, one finds for both nonlinear
Compton and Breit-Wheeler processes that $\theta_{\infty} = 3\pi
c_{\eps}\xi^{2}\Phi/2$, where $c_\eps = 1$ ($c_\eps = 1/2$) for a circularly
(linearly) polarised background. This finally leads to
\bea
I_2 = 4\pi\Phi\sin X&&\left[\cos X\, \trm{Ci}\,Y-\sin X\, \trm{Si}\,Y  \right. \nn \\
&& \left. + \frac{\pi}{2}\sin X - \frac{1}{Y} \sin \left(X+Y\right)\right],
\eea
where $X = \theta_{\infty}g(s)/2$ and $Y=4\pi\Phi g(s)$. This is related to
recent studies of interference effects in a double-pulse background
\cite{ilderton19,Ilderton:2020dhs}.

The remaining integral, $I_3$, collects the contributions where the average
phase $\phi$ is outside the pulse, while the phase difference $\theta$ is large
enough that $\phi-\theta/2$ reaches back into the pulse,
\bea
I_{3} = \int_{2\pi\Phi}^{\infty} \ud\phi
\int_{2(\phi-2\pi\Phi)}^{2(\phi+2\pi\Phi)} \frac{\ud\theta}{\theta}& & \left\{ -\sin\left[g(s)\theta\right] \right. \nn  \\ && \left. + \sin\left[g(s)\mu(\phi,\theta)\right]\right\} ,\nn
\eea
The integrand oscillates with a slowly decaying amplitude for $\phi > 2\pi\Phi$
outside the pulse. As the oscillations are regular, they can be handled by using
many data points. We also expect (and will show later) that contributions from
outside the pulse are important mainly in the infra-red region of the spectrum,
where we have an analytical expression for the limit.

Finally, $I_{4}$ is just the evaluation of the full integral in
\eqnref{eqn:pulse1}, for $\phi \in [0,2\pi\Phi]$, $\theta \in [0, 2(2\pi\Phi +
\phi)]$, i.e. ``on top of'' the pulse. As this is a well-defined, finite
integration range, convergence can be assured by simply increasing the sampling
resolution of the integrand.
\begin{figure}[!!h]
\centering
\includegraphics[width=0.99\linewidth]{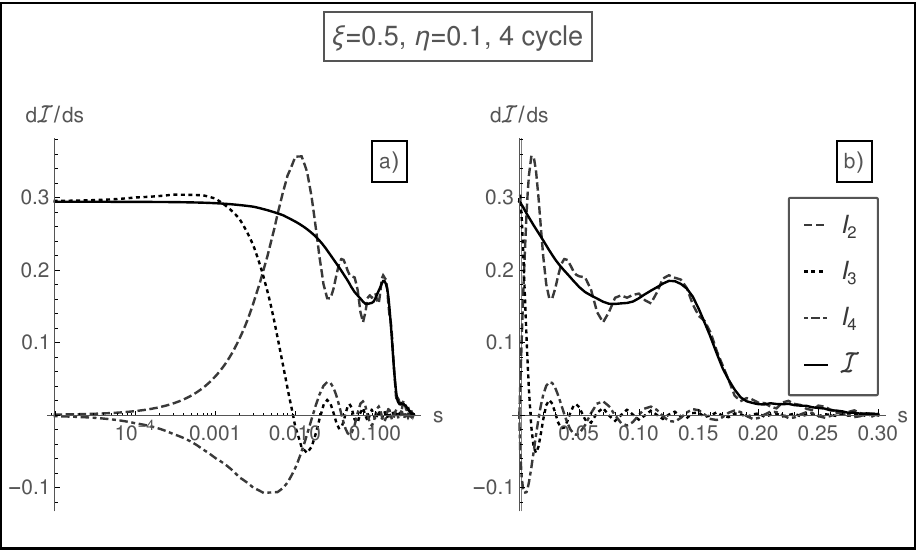}
\caption{A demonstrative plot showing how different parts of the integration
    region contribute to the spectrum (here, for a linearly polarised pulse)
    using a) a log scale and b) a linear scale.} \label{fig:decompPlot1}
\end{figure}

The contribution of each part of the phase integration plane $(\phi,\theta)$ to
the spectrum is shown, for example parameters, in Fig.~\ref{fig:decompPlot1}.
This demonstrates that in the infra-red limit, $s \to 0$, the integral $\mcI$
from (\ref{eqn:pulse1}) is dominated by the sub-integral $I_2$, i.e.\ by
contributions from phase regions located \emph{outside} the pulse.  On the one
hand, this agrees with intuition based on the uncertainty principle---the lowest
photon energies require the longest interaction of the electron with the
background as has already been pointed out in the literature for nonlinear
Compton  scattering~\cite{king15d}. On the other hand, when studying the
infra-red, one should take into account soft contributions from higher-order
processes~\cite{dinu12}.

%
\section{Nonlinear Compton scattering \label{sec:NLC}}
%

\subsection{Circularly polarised plane wave}\label{sec:NLCcirc}
\begin{figure}[h!!]
\centering
\includegraphics[width=0.99\linewidth]{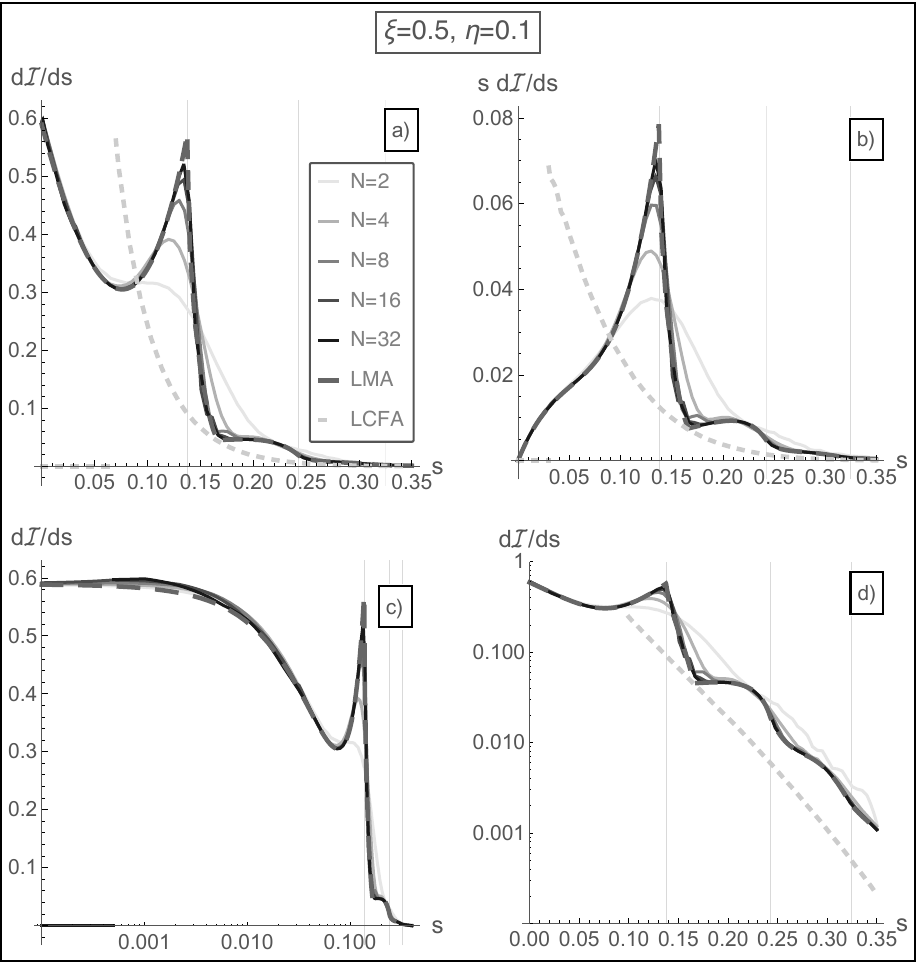}
\caption{The photon spectrum from nonlinear Compton scattering in a
    \emph{circularly} polarised background, in the high-energy, weakly nonlinear
    regime, normalised by $N/2$ for pulses with different numbers of cycles, $N$. The
    locally-constant field approximation (light short-dashed line) poorly approximates
    the spectrum, whereas the LMA (dark long-dashed line) captures the harmonic
    structure and becomes more accurate as the length of the pulse increases.
    Plotted left-to-right is: a) the yield spectrum; b) the energy spectrum; c)
    the IR part of the spectrum (log-linear); d) the UV part of the spectrum
    (log). The vertical solid lines here and in the following figures
    correspond to the positions of the harmonic edges calculated for an
    infinite monochromatic plane wave.}\label{fig:NLCCP1}
\end{figure}

\begin{figure}[h!!]
\centering
\includegraphics[width=0.99\linewidth]{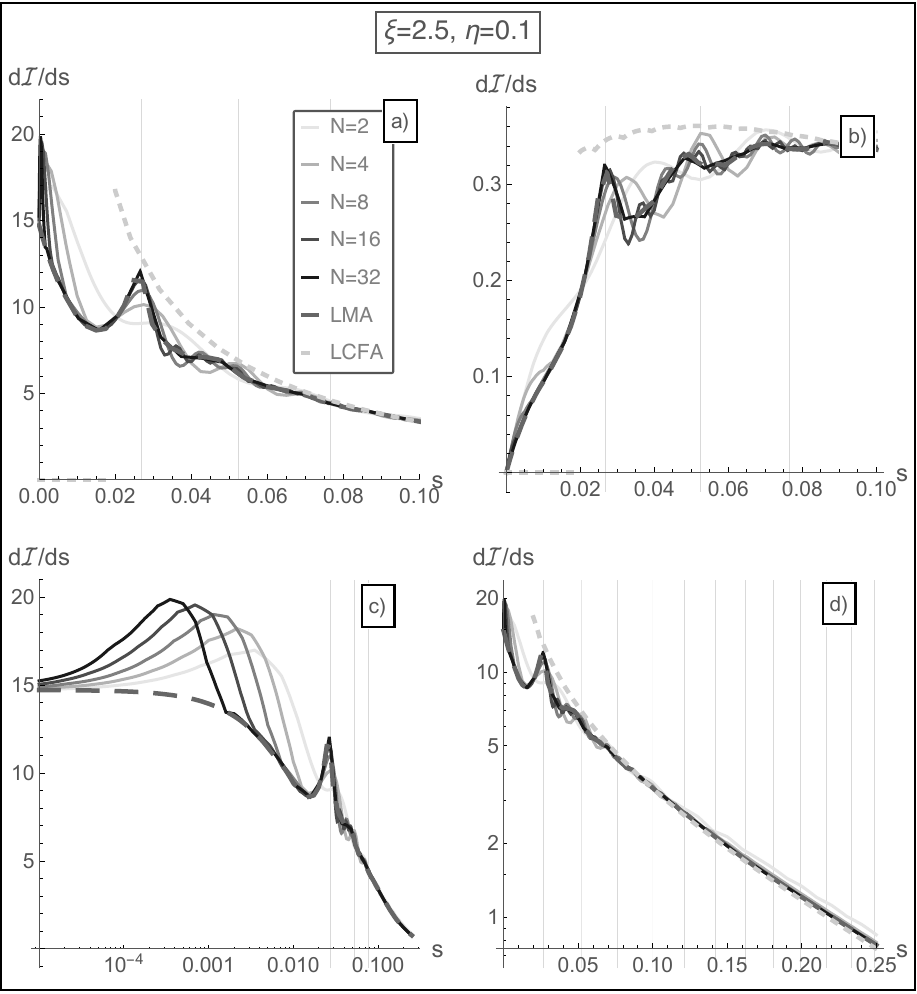}
\caption{The photon spectrum from nonlinear Compton scattering in a \emph{circularly}-polarised background, in the high-energy nonlinear regime, normalised by half the number of laser cycles, $N/2$. The locally-constant field approximation (light short-dashed line) approximates the spectrum well for values of $s$ corresponding to higher harmonics. The LMA (dark long-dashed line) captures both the harmonic structure and the large-$s$ behaviour and becomes more accurate as the length of the pulse increases. Plotted left-to-right is: a) the yield
spectrum; b) the energy spectrum; c) the IR part of the spectrum (log-linear);
d) the UV part of the spectrum (log). }\label{fig:NLCCP2}
\end{figure}

Having evaluated the full QED integrals for a pulse, we  can now compare with  the LMA. The latter is numerically more efficient, but also implies enhanced analytical control as it typically results in well-known special functions. Beyond these immediate advantages, our 
motivation to improve standard literature approximations is three-fold: (i)
to have a locally defined rate which could in principle be implemented in numerical simulation
codes; (ii) to be able to resolve the harmonic structures present in the exact
QED probabilities with this approximation; and (iii) to be able to work in the
moderate intensity regime, $\xi \sim 1$, relevant for current state-of-the-art
laser facilities. By construction, item (i) is readily provided by the LMA. To
test the LMA for the other two goals, we will benchmark it against numerically
integrated exact QED probabilities, beginning with the process of nonlinear
Compton scattering. 

Consider the interaction of an electron, initial invariant energy parameter
$\eta_{e} = k \cdot p/m^{2}$, with the plane wave
\begin{align}\label{eqn:CircPot}
	a_{\mu}(\phi)
	=
	m \xi
	\cos^2\Big(
		\frac{\phi}{\Phi}	
	\Big)
	\big(
		\varepsilon_{\mu}
		\cos\phi
		+
		\bar{\varepsilon}_{\mu} 
		\sin\phi
	\big) 
	\;,
\end{align}
which has circular polarisation and envelope $f\sim \cos^{2}$. The LMA to the
nonlinear Compton spectrum in this setup is given in (\ref{eqn:NLCcirc}). In
Fig.~\ref{fig:NLCCP1} we compare the photon spectrum predicted by the LMA with
the exact QED result, for the parameters $\xi = 0.5$ and  $\eta_{e} = 0.1$, and
various pulse lengths $\Phi$. This is  the low intensity, high-energy regime
which will be probed at, for example, LUXE~\cite{Abramowicz:2019gvx}. In this
regime, the locally-constant field approximation, valid for $\xi^2/\eta_{e} \gg
1$~\cite{Khok1}, is no longer applicable and fails by a large margin as
demonstrated in Fig.~\ref{fig:NLCCP1}.

Each of the plots (a)--(d) in Fig.~\ref{fig:NLCCP1} shows the spectra for the
LMA (dark long-dashed line), the locally-constant field approximation (light short-dashed line) and
the numerically integrated exact QED results, the latter of which is
plotted for various pulse lengths. (We recall the number of cycles $N$ and the pulse
duration $\Phi$ are related by $\Phi = 2 N$.) The numerically integrated exact
QED spectra have been normalised by $N/2$ to facilitate comparison. As discussed
above, one of the steps in deriving the LMA for a given process is to first
apply the slowly-varying envelope approximation, which assumes that the pulse
duration is sufficiently long such that derivatives of the profile can be
neglected. We can see the consequences of this in Fig.~\ref{fig:NLCCP1}. As the
pulse duration is increased, the LMA result remains the same (when normalised by pulse duration), but the results from the numerical
integration of the exact QED probability become progressively more peaked around
the first harmonic, and the agreement between this and the LMA improves. In all
cases, the locally-constant field approximation completely misses not only the
key harmonic structures and the infra-red limit, but also fails in the
high-energy, $s \to 1$, regime. This is characteristic of the locally-constant
field approximation for $\xi < 1$. 
 
Fig.~\ref{fig:NLCCP2} we show the same spectra as before, however for the
increased field strength of $\xi = 2.5$. We are now in a regime where the
locally-constant field approximation is able to more accurately capture at least
the $s \to 1$ behaviour of the spectra, but we can see that the LMA is still
vastly superior. In fact, in Fig.~\ref{fig:NLCCP2}c we can distinguish three
distinct regions of the spectrum on the interval $0 < s < 1$, defined in
relation to the position of the first harmonic/Compton edge, which for a
monochromatic plane wave is located at  $s_1 = 2 \eta_{e} / (1 + \xi^2 +
2\eta_{e})$.  There is the far infra-red sector where $0 < s \ll s_1$, the
harmonic range where $s > s_1$ which includes all of the harmonic structure of
the spectrum, and the intermediate regime where $s \lesssim s_1$. In both the
far infra-red and the harmonic range the LMA gives a very good agreement with
the numerically integrated exact QED spectrum, outperforming the
locally-constant field approximation in both cases. One of the most striking
improvements in this regard is the agreement between the LMA and the exact QED
spectrum in the far infra-red, $s \to 0$ limit. This agreement is not only
visible numerically; one can trivially derive the correct $s \to 0$ limit from
the LMA, as shown in Appendix~\ref{app:Smalls} where we also provide a novel
derivation of the limit from the exact QED probability.

The second area in which the LMA performs well is in the harmonic range.  For
sufficiently long pulses, which in Fig.~\ref{fig:NLCCP2} corresponds to $8$
cycles (full-width-half-max duration of around $11\,\trm{fs}$ for a
$800\,\trm{nm}$ carrier wavelength), the LMA not only predicts the correct
position of the leading harmonic in the spectrum, but is also accurate in
predicting the locations and magnitudes of the sub-leading harmonics. 

The only part of the spectrum in which the LMA deviates somewhat from the exact
QED result is the intermediate regime where $s \lesssim s_1$. It turns out that
this sector of the spectrum contains features which, to the best of our
knowledge, have not been extensively commented on in the literature. Most
numerical investigations of the exact QED spectrum/probability are compared to
the locally-constant field approximation, which is well known (i) to not capture
harmonic structure and (ii) to diverge towards the infra-red.  The LMA, however,
yields the correct infra-red limit, $s \to 0$,  and very good agreement in the
harmonic range, but does not capture the full structure of the spectrum in the
intermediate range. In each of the spectra coming from the numerically
integrated exact QED results there is a clear ``bump'' in the range just before
the first harmonic. This same feature can be seen in various other works in the
literature, see for example
\cite{Ilderton:2018nws,DiPiazza:2018bfu,Blackburn:2018sfn}. 

A qualitative explanation for these additional peaks is that a pulse profile  introduces additional frequency scales in the dynamics, analogous to the usual harmonics found at locations determined by the carrier frequency scale of the background, see  e.g.~\cite{Boca:2009zz,Seipt:2010ya,Mackenroth:2010jr}. For the current choice of a $\cos^{2}$ pulse envelope, we found that the approximate position of these peaks can be determined as follows. One first introduces a rescaled frequency, $\tilde{k}^{0} = k^{0}/2I $, where $k^{0}$ is the carrier wave frequency and $I$ is the integral\footnote{For circular polarisation, i.e.\ the choice (\ref{eqn:CircPot}), one finds $I = \pi\Phi/2$. An analogous argument for linear polarisation (see below) employs the scaling $\tilde{k}^{0} = k^{0}/\sqrt{2} \pi \Phi$.} of the pulse profile, $f$. One then calculates the position of the first harmonic/Compton edge, $s_{1}$, using the rescaled energy parameter $\eta_{e}\to \eta_{e}/2I$. As pulse duration increases, the additional broad peaks get pushed further back into the infra-red and are smoothed out, eventually disappearing in the infinite plane wave (monochromatic) limit. Therefore an improvement of the accuracy of the LMA in this part of the spectrum might be achieved by including higher order terms in $1/\Phi$, i.e.\ the slowly-varying-envelope part of the approximation. The amplitude of these peaks also decreases significantly as $\xi$ falls below unity.

Fig.~\ref{fig:NLCCP2} also shows that in the UV range, $s\to1$, there is good
agreement between the LMA and the locally constant field approximation for
$\xi=2.5$. However, this is no longer true when $\xi=0.5$ as in
Fig.~\ref{fig:NLCCP1}. To capture the UV limit in more detail one could adopt
the methods of  \cite{torgrimsson18, torgrimsson19} and use the saddle point
method, noting that, in the exponent, the pre-factor of the Kibble mass is
proportional to $ (1-s)^{-1}$. Following this route, though, is beyond the scope
of our present discusion.

The case of a circularly polarised plane wave pulse gives the simplest form of
the LMA due to the additional symmetries of the choice of background. The
approach can, however, still be used for the case of linear polarisation, to
which we now turn.

\begin{figure}[t!!]
\centering
\includegraphics[width=0.99\linewidth]{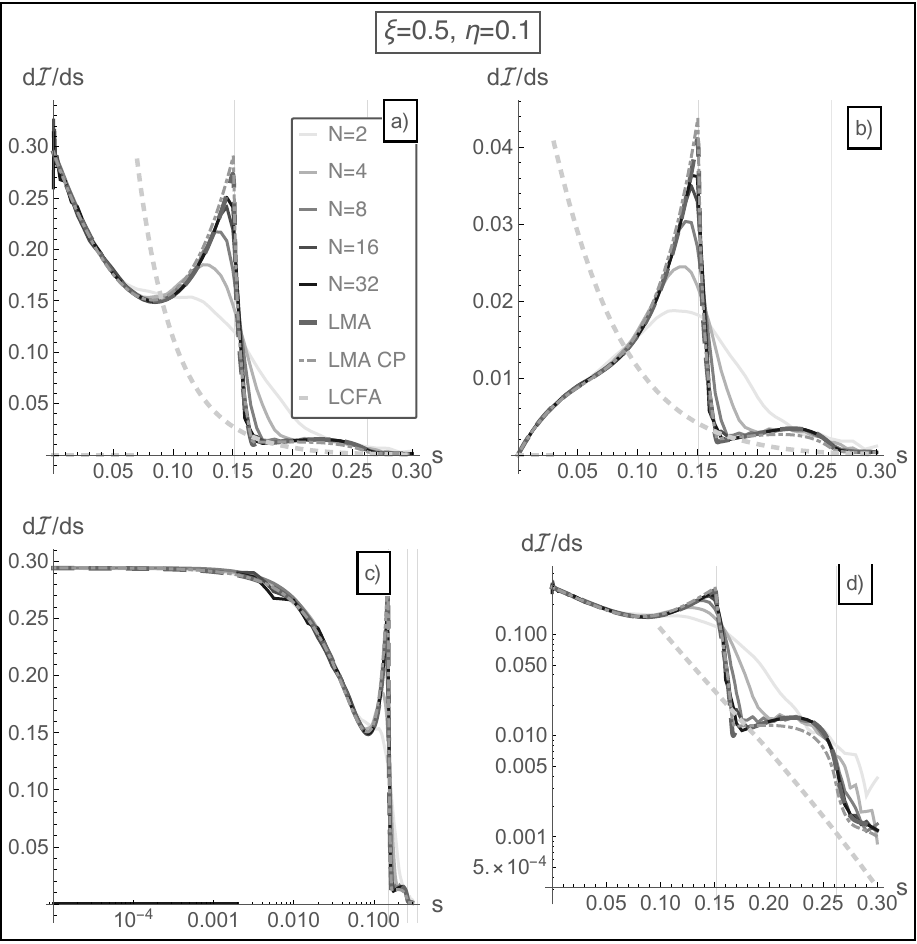}
\caption{The photon spectrum from nonlinear Compton scattering in a \emph{linearly} polarised background, in the high-energy, weakly nonlinear regime, normalised by half the number of laser cycles, $N/2$. The agreement of the LMA (dark long-dashed line) and disagreement of the locally-constant field approximation (light short-dashed line) with the numerically exact results is similar to the circularly polarised case. The dot-dashed line is the spectrum acquired by taking the LMA for a \emph{circularly} polarised background and rescaling the intensity parameter $\xi\to\xi/\sqrt{2}$.}\label{fig:NLCLP1}
\end{figure}
\begin{figure}[b!!]
\centering
\includegraphics[width=0.99\linewidth]{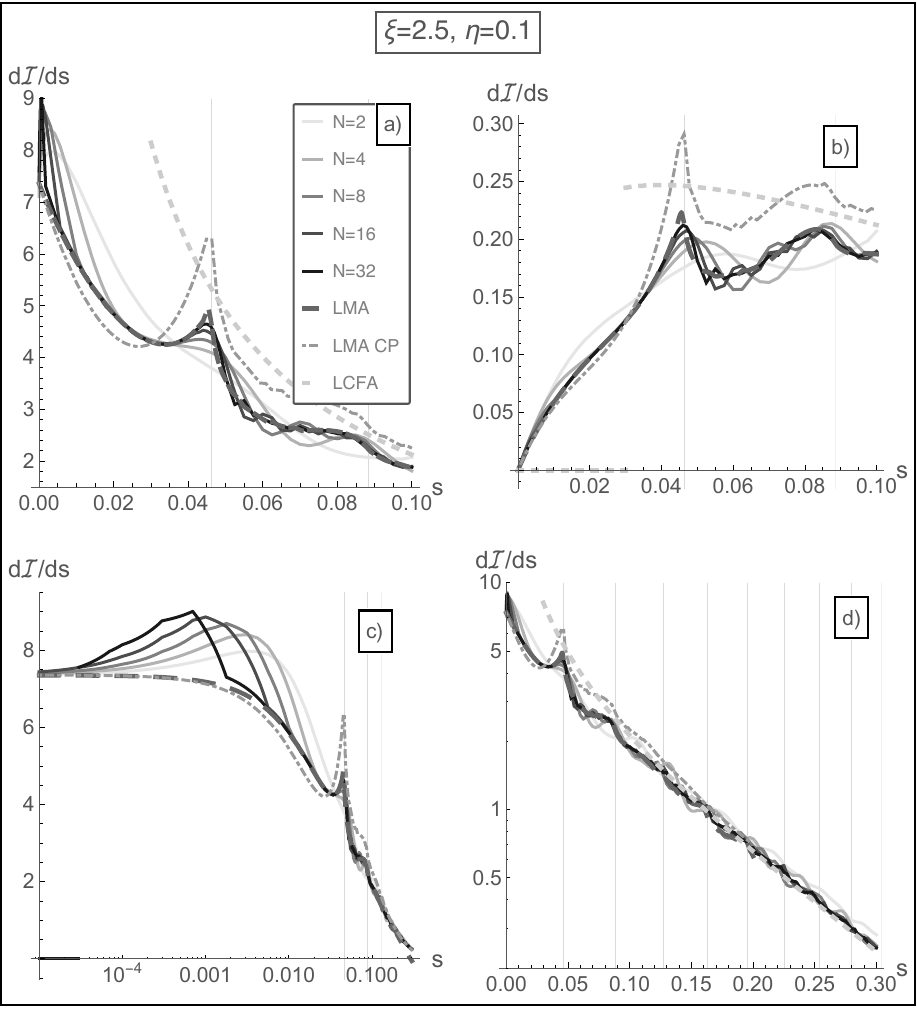}
\caption{The photon spectrum from nonlinear Compton scattering in a \emph{linearly} polarised background,
in the high-energy, nonlinear regime, normalised by half the number of laser cycles. The agreement of the LMA (dark long-dashed line) and the locally-constant field approximation (light short-dashed
line) with the exact pulsed results is similar to the circularly polarised case. The dot-dashed line is
the spectrum acquired by taking the LMA for a \emph{circularly} polarised
background and replacing the intensity parameter $\xi\to\xi/\sqrt{2}$. Unlike in
the weak-field regime, the linearly polarised LMA is not well approximated
by rescaling the intensity parameter in the circularly polarised LMA.}\label{fig:NLCLP2}
\end{figure}

\subsection{Linearly polarised plane wave \label{sec:NLClin}}

As above, we compare the LMA for a linearly polarised background field with the
numerically integrated exact result for a fixed electron energy $\eta_{e} = 0.1$
and field strengths $\xi = 0.5$ (Fig.~\ref{fig:NLCLP1}) and $\xi = 2.5$
(Fig.~\ref{fig:NLCLP2}). In this case the LMA is given by (\ref{eqn:NLClin}).
Even for infinite monochromatic plane wave fields, the probability of nonlinear
Compton scattering for a linearly polarised background field has extra structure
compared to the circularly polarised case. The same is true for the LMA in a
pulsed linearly polarised field. The source of the extra structure is that for
linear polarisation the term which is quadratic in the background field in the
classical action (\ref{def:ClassAct}) is dependent on both the slow oscillations
due to the pulse profile, and the fast oscillations of the carrier frequency of
the plane wave. Within the LMA, this results in a non-trivial integration over
the angular spread of the emitted photons.  As a consequence (see
Appendix~\ref{app:NLCLMA} for details), there remains a double harmonic sum,
compared to the circularly polarised case, where it simplifies due to the extra
symmetry in the background.  Hence, it is not possible to simply take the
textbook expression for linearly polarised monochromatic plane waves
\cite{Berestetsky:1982aq} and localise the field intensity, $\xi \to \xi f$, as
could be done in the circularly polarised case.  In principle, the additional
structure of a double-harmonic sum allows for the possibility of interference
effects between the harmonics. However,  in the intermediate intensity,
high-energy regime, we did not find any appreciable contribution from this
interference.

For weak fields, $\xi < 1$, the low-energy part of the spectrum, i.e.\ the
region $s \lesssim s_{1}$ below the first harmonic, $s_{1}$,  is well
approximated by the perturbative contribution from the squared potential,
$a^{2}$.  In this case, the linearly polarised LMA turns out to be
well-approximated by taking the circularly polarised LMA and making the
replacement $\xi\to\xi/\sqrt{2}$, as is demonstrated in (\ref{fig:NLCLP2}).
Because of this, rescaling the circularly polarised result is a method which has
been used to implement rates for linear polarisation in numerical codes.

However, this method fails for $\xi>1$. In this regime, higher harmonics,
proportional to $a^{2n}$ for the $n$th harmonic, contribute to the spectrum and
can no longer be obtained through a simple modification of the circularly
polarised LMA. This impact of the background polarisation at higher values of
the field strength is demonstrated in \figref{fig:NLCLP2}. Although the position
of the harmonics is still correctly predicted by the rescaled circularly
polarised LMA, their amplitude is not, nor is the overall shape of the spectrum
correctly captured: the rescaled circularly polarised result gives an
underestimate for the smallest values of $s$, but an overestimate for larger
values. Hence, the linearly polarised LMA proper, rather than the rescaled
circularly polarised LMA, must be used in the intensity regime of upcoming
experiments \cite{Abramowicz:2019gvx}.

From both the circular and linear polarisation cases just discussed one notes
that the higher the field strength $\xi$, the better the agreement between LMA
and locally-constant field approximation in the ultra-violet (large-$s$) regime.
In appendix~\ref{app:LCFA} we show explicitly that this is not just some
numerical accident. Indeed, we will derive the locally-constant field
approximation as the high-field limit of the LMA.

%
\section{Nonlinear Breit-Wheeler \label{sec:BW}}
%

So far our focus has been on implementing and analysing the LMA for nonlinear
Compton scattering. In principle, however, the LMA can be applied to any QED
scattering process in a plane wave background. As another example, consider
nonlinear Breit-Wheeler pair production, where an initial photon decays into an
electron-positron pair. The derivation of the LMA for this process follows the
same route as for nonlinear Compton scattering (see appendix~\ref{app:NLCLMA}),
and we again find that in the case of a circularly polarised plane wave the
final differential probability is simply the textbook result in a monochromatic
plane wave \cite{Berestetsky:1982aq} with a localisation of the field strength,
$\xi \rightarrow \xi f$, see (\ref{eqn:LMABW}) in the appendix. A well known
feature of the nonlinear Breit-Wheeler process is the strict lower bound,
$n_\star$, on the harmonic number contributing for a given field strength and
initial photon energy. This is because the outgoing particle states are massive,
so that their production can only proceed above an energy threshold.

For a monochromatic plane wave, the lower bound is given by
$n_{\star}^{\text{mono}} = 2 (1 + \xi^2)/\eta_{\gamma}$, where
$\eta_{\gamma}=k\cdot \ell/m^2$ is the energy parameter for the incident photon
with four-momentum $\ell$. Comparing this to (\ref{eqn:BWparams}), we can see
that for a pulse there are points along the phase for which the minimum harmonic
$n_{\star} < n_{\star}^{\text{mono}}$ for the same $\xi$ and $\eta_{\gamma}$. 

At first glance, this would appear to mean that at the wings of the pulse, as $f
\to 0$, the minimum harmonic contributing would decrease, and since Bessel
harmonics of lower order are typically greater in magnitude, that the process
would be more probable at lower field strengths. One has to keep in
mind, though,  that the argument $z(\phi)$ of the Bessel function depends on the pulse profile $f$ and vanishes in the limit $f \to 0$. The only Bessel function surviving this limit is $J_{0}$. However, since the harmonic sum in
(\ref{eqn:LMABW}) is over strictly positive $n > 0$ and thus excludes $J_0$, there is no contribution to the probability for $f \to 0$. Hence, in comparison to nonlinear Compton scattering, the nonlinear Breit-Wheeler process will still require either very high field strengths, for which the locally-constant field approximation
should be a good approximation, or very high initial photon energies.

\begin{figure}[t!!]
\includegraphics[width=0.7\linewidth]{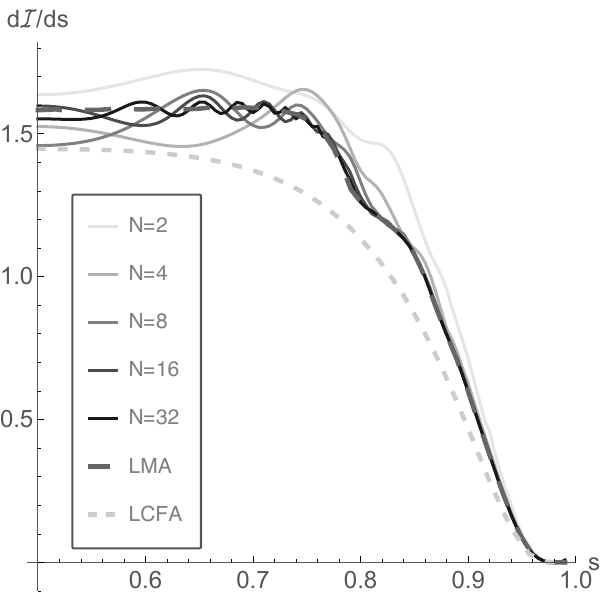}
\caption{The spectrum of electrons produced in nonlinear Breit-Wheeler pair production for
$\xi=1$, $\eta_{\gamma}=3$. A comparison of the locally-constant field approximation (light short-dashed line) and the LMA (dark long-dashed line) with the exact numerical result in pulses of different numbers of cycles, $N$.}\label{fig:NBW1}
\end{figure}

For both the Compton and Breit-Wheeler processes, the momentum taken from the field increases with field strength, and the harmonic structure becomes less
well defined. In order to demonstrate the LMA for the  Breit-Wheeler process,
the centre-of-mass energy should be close to the pair rest-energy in order that
only very few laser photons are required for pair production to take place. In
Fig.~\ref{fig:NBW1}, we demonstrate such a situation, where we present the
spectrum of electrons produced by a head-on collision of a $250\,\trm{GeV}$
photon ($\eta_{\gamma} = 3$)  with a laser pulse of intensity $\xi =1$. We note
that the harmonic structure of the spectrum for long pulses is well-approximated
by the LMA, whereas the locally-constant field approximation both misses the
harmonics in the spectrum and under-predicts the yield.
		
%
\section{Summary \label{sec:Summary}}
%

Motivated by the need to improve the theoretical tools required for supporting
state-of-the-art laser experiments probing the high-intensity regime of QED, we
have introduced here the locally monochromatic approximation (LMA). 

This technique treats the quickly- and slowly-oscillating components of laser
field profiles differently, in order to improve on the accuracy of the existing
locally-constant field approximation, which essentially treats all field
components as slowly varying. Oscillations due to the carrier frequency of the
laser field are treated exactly, while the slowly-varying field envelope degrees
of freedom are treated in a local expansion. Therefore, the accuracy of the LMA increases with increasing pulse duration as we have shown by  comparing directly with exact QED results. This conclusion agrees with other works that have compared a train of monochromatic pulses with single short-pulse spectra \cite{krajewska12a,krajewska12b}. Although we have not included it here, the LMA could easily be extended to include a carrier-envelope-phase, since the separation between fast and slow time scales would remain (see e.g. \cite{Titov:2019kdk} for an example of this applied to the slowly-varying-envelope approximation).

The LMA (or its precursors) have been used in several numerical codes, albeit in an ad-hoc fashion. To put the LMA on a firmer basis, we provide the first derivation from, and the first benchmarking against, QED in a plane wave background. In doing so we have identified the character of expansions at work and established how the accuracy improves with pulse duration. Finally, we have located spectral features in the mid-infra-red that are missed by this approximation.

We note, however, that despite being local in the phase variable, the LMA is unsuitable for intense laser-matter interactions where plasma is present. This is because the LMA relies on the presence of structures
particular to laser fields, essentially a central frequency and an envelope,
which normally are absent in a plasma. Instead, the LMA can be thought to extend the LCFA up to higher energies and down to intermediate and low intensities, in situations where the background field is a laser pulse of well-characterised shape. Such a situation is to be found in upcoming high-energy experiments~\cite{Abramowicz:2019gvx}.
When applicable, the LMA correctly resolves harmonic structure in particle spectra. Whilst it is known that these can be washed out due to multi-particle effects \cite{angioi16}, they have been observed in experiments
~\cite{Bamber:1999zt,babzien06,sakai15,Khrennikov15}. The washing-out effect is expected to be less significant if the electron beam has a narrow momentum spread. A further advantage of the LMA is its capability to capture the infra-red limit of nonlinear Compton
scattering. In contrast, the locally-constant field approximation is well-known to fail in this regard.

In this paper we have considered the first-order processes of nonlinear Compton
scattering and nonlinear Breit-Wheeler pair production, but the LMA could also
be extended to higher-order processes such as trident pair production (see
e.g.~\cite{Ritus:1972nf,Ilderton:2010wr,King:2018ibi,Mackenroth:2018smh,Dinu:2019wdw,Acosta:2019bvh})
and double nonlinear Compton scattering (see
e.g.~\cite{Morozov:1975uah,Lotstedt:2009zz,Seipt:2012tn,Mackenroth:2012rb,King:2014wfa,Dinu:2018efz}). This extension is not trivial, as it would need to deal with the appearance of resonant singularities in dressed propagators \cite{oleinik67,krajewska11} and is therefore a subject for further work.

\acknowledgements

The authors thank Anton Ilderton for many useful discussions and a careful reading of the manuscript. B.K.\ and A.J.M.\  are supported by the EPSRC grant EP/S010319/1.
\appendix
\section{Detailed derivation of the LMA \label{app:NLCLMA}}

In Sec.~\ref{sec:LMA} we presented only the key steps involved in calculating
the LMA for a high-intensity QED process. To be more explicit, we turn to an
example derivation for the process of nonlinear Compton scattering in a plane
wave pulse. We will give a thorough account of the calculation for a circularly
polarised pulse, and provide details about the technical differences in the
linearly polarised case. We note that the calculation of the LMA for nonlinear
Breit-Wheeler pair production discussed in Sec.~\ref{sec:BW} follows a
completely analogous procedure. Hence, it will be sufficient to simply quote the
final result below. Once the slowly-varying envelope approximation has been
applied to the exponent of the scattering amplitude, the remaining analysis
amounts to a generalisation of the calculation in a purely monochromatic plane
wave (see for example~\cite{Berestetsky:1982aq} for the circularly polarised
case).

Nonlinear Compton scattering is the process by which an electron of 4-momentum
$p$ scatters off a background plane wave pulse to emit a photon of momentum
$\lp$ and polarisation $\epsilon_{\lp}^*$, $e^-(p) \rightarrow e^-(\pp) +
\gamma(\lp)$. The amplitude is given by the standard $S$-matrix element
\begin{align}
S_{\text{NLC}}
=
-
i
e
\int
\ud^4 x\,
\bar{\Psi}_\pp(x)
\slashed{\epsilon}^*_\lp
\Psi_p(x)
e^{i \lp \cdot x}
.
\end{align}

The explicit representation of the Volkov wavefunctions (\ref{def:Volkov}) and
some  trivial integrations lead to the representation (\ref{def:Amplitude}) with
reduced amplitude
\begin{align}\label{eqn:Reduced}
\mcM_{\text{NLC}}
=&
-
i
e
\int
\!
\ud\varphi
\;
\mcS(\varphi)
\exp\left(
i 
\int^{\varphi}_{-\infty}
\frac{\lp \cdot \pi_p}{k \cdot (p - \lp)}
\right)
\;.
\end{align}
The integrand involves a spin structure
\begin{align}
\mcS(\varphi)
=
\bar{u}_{\pp}
\bigg(
1
+
\frac{\slashed{a}(\varphi) \slashed{k}}{2 k \cdot \pp}
\bigg)
\slashed{\epsilon}^*_\lp
\bigg(
1
+
\frac{\slashed{k} \slashed{a}(\varphi)}{2 k \cdot p}
\bigg)
u_{p}
\;.
\end{align}
and an exponential given in terms of the kinetic momentum of a classical
electron in a plane wave background,
\begin{align}\label{def:Kinetic}
\pi_p^\mu(\varphi)
=
p^\mu
-
a^\mu(\varphi)
+
\frac{2 p \cdot a(\varphi) - a^2(\varphi)}{2 k \cdot p}
k^\mu
\;.
\end{align}
We proceed now to the particular case of a circularly polarised pulse.

\subsection{Circularly polarised plane wave pulse}

The circularly polarised plane wave pulse is given by (\ref{def:Gauge}) with
$\delta = \pi/4$ and a normalisation factor of $\sqrt{2}$ such that
$\mathsf{max} \, (\left|a^\mu/m\xi\right|) = 1$, 
\begin{align}\label{def:GaugeCirc}
a^\mu(\varphi)
=&
m
\xi  \,
f\Big(\frac{\varphi}{\Phi}\Big)
\big(
\varepsilon^\mu
\cos\varphi
+
\bar{\varepsilon}^\mu
\sin\varphi
\big)
\;.
\end{align}
The term quadratic in the gauge potential in (\ref{def:Kinetic}) only contains
the slow timescale in $\varphi$:
\begin{align}\label{eqn:GaugeSqCirc}
a^2(\varphi)
=
-
m^2 
\xi^2
f^2\Big(\frac{\varphi}{\Phi}\Big)
\;,
\end{align}
and so the slowly-varying envelope approximation  (\ref{eqn:LinearIntApprox})
need only be applied to the other terms linear in $a^\mu$. This results in
\begin{align}\label{eqn:Exponent}
\int^{\varphi}_{-\infty}
\frac{\lp \cdot \pi_p}{k \cdot (p - \lp)}
\simeq
G(\varphi)
-
\alpha_c\Big(\frac{\varphi}{\Phi}\Big)
\sin\varphi
+
\alpha_s\Big(\frac{\varphi}{\Phi}\Big)
\cos\varphi.
\end{align}
The function $G(\varphi)$ is slowly varying with $\varphi$,
\begin{align}\label{def:G}
G(\varphi)
=
\frac{s}{2 \eta_{e} (1 - s)}
&\bigg[
\bigg(
1
+
\frac{|\bm{\ell}_\LCperp^\prime - s\bm{p}_{\LCperp}|^2}{s^2 m^2}
\bigg)
\varphi
\nonumber\\
&
\qquad
+
\int^{\varphi}_{-\infty}
\ud\psi\;
\xi^2
f^2\Big(\frac{\psi}{\Phi}\Big)
\bigg]
\;,
\end{align}
and depends on $\eta_{e} = k \cdot p / m^2$, the normalised measure of the
electron's light-front momentum, and $s = k \cdot \lp / k \cdot p$, the
light-front momentum fraction of the outgoing photon. The rapidly oscillating
terms $\{\cos\varphi,\sin\varphi\}$ each have a slowly-varying pre-factor,
\begin{align}
\alpha_c\Big(\frac{\varphi}{\Phi}\Big)
=&
\frac{\xi f(\frac{\varphi}{\Phi})
}{\eta_{e} m (1 - s)}
(\lp - sp) \cdot \varepsilon
\;,
\nonumber\\
\alpha_s\Big(\frac{\varphi}{\Phi}\Big)
=&
\frac{\xi f(\frac{\varphi}{\Phi})
}{\eta_{e} m (1 - s)}
(\lp - sp) \cdot \bar{\varepsilon}
\;,
\label{def:Alphas}
\end{align}
respectively, and depend on a 4-vector $\mcL = \ell^\prime - s p$, projected
onto the polarisation directions, $\varepsilon$ and $\bar{\varepsilon}$, of the
background. Defining an angle $\vartheta$, 
\begin{align}
\vartheta
=
\tan^{{-1}}
\frac{\alpha_s}{\alpha_c}
\;,
\end{align}
allows us to write
\begin{align}
\alpha_c\Big(\frac{\varphi}{\Phi}\Big)
=&
z\left(\frac{\varphi}{\Phi}\right) 
\cos\vartheta
\;,
\quad&
\alpha_s\Big(\frac{\varphi}{\Phi}\Big)
=&
z\left(\frac{\varphi}{\Phi}\right) 
\sin\vartheta
\;,
\end{align}
(we will use the shorthand $z(\vphi)\equiv z(\vphi/\Phi)$ in what follows) such that
\begin{align}
z(\varphi)
=
\sqrt{\alpha_c^2 + \alpha_s^2}
=
\sqrt{
	\frac{\xi^2 f^2(\frac{\vphi}{\Phi})}{\eta_{e}^2 m^2 \; (1 - s)^2}
	|\bm{\lp} - s\bm{p}|^2
}
\;.
\end{align}
This drastically simplifies the exponent (\ref{eqn:Exponent}), and the reduced
amplitude (\ref{eqn:Reduced}), which becomes
\begin{align}\label{eqn:ReducedCirc}
\mcM_{\text{NLC}}
=&
-
i
e
\int
\!
\ud\varphi
\;
\mcS(\varphi) \, 
e^{ i \{
G(\varphi)
-
z(\varphi)
\sin (\varphi - \vartheta)
\}}
.
\end{align}

The probability is now calculated in the usual way by averaging/summing over
incoming/outgoing spins and polarisations and integrating over the outgoing
particle phase space with the result
\begin{widetext}
\begin{align}\label{eqn:NLC}
\bbP_{\text{NLC}}^{\text{(circ)}}
=&
\frac{\alpha}{16 \pi^2 (k \cdot p)^2}
\!
\int
\!
\ud\varphi
\!
\int
\!
\ud\varphi^\prime
\!
\int
\!
\!
\!
\frac{\ud s}{s (1 - s)} 
\!
\int
\!
\ud|\bm{\mcL}_{\perp}|^2
\int
\!
\ud \vartheta
\;
\mcT_{\text{NLC}}(\varphi,\varphi^\prime)
\nonumber\\
&
\times
\exp\left[
i 
G(\varphi)
-
i 
G(\varphi^\prime)
-
i
z(\varphi)
\sin\big(\varphi - \vartheta)
+
i
z(\varphi^\prime)
\sin\big(\varphi^\prime - \vartheta)
\right]\,.
\end{align}
\end{widetext}
Here, we have introduced the fine structure constant $\alpha$ and the auxiliary quantity
\begin{align}\label{eqn:Trace}
\mcT_{\text{NLC}}(\varphi,\varphi^\prime)
=&
-
2
m^2
+
\bigg(
1
+
\frac{s^2}{2(1 - s)}
\bigg)
\big(a(\varphi) - a(\varphi^\prime)\big)^2
,
\end{align}
(up to a factor) representing the trace, $\tr \bar{\mcS} \mcS$, from the spin
sum/average.  For the perpendicular photon momentum integrals one has $\ud^2
\bm{\ell}^\prime_{\perp} = \ud^2 \bm{\mcL}_\perp$ or, in polar coordinates,
\begin{align}
\int
\!
\ud^2 \bm{\ell}^\prime_{\perp}
=
\half
\int
\!
\ud |\bm{\mcL}_\perp|^2
\int
\!
\ud \vartheta
\; .
\end{align}

The trace term (\ref{eqn:Trace}) also depends on the gauge potential. However,
the rapidly oscillating parts of both the exponential and the pre-exponential
can always be combined into the Bessel generating function
(\ref{def:Generating}).  Doing so, we expand each term in the probability into
sums over Bessel harmonics, writing, in this expression, $z\equiv z(\varphi)$
and $z'\equiv z(\varphi')$, 
\begin{widetext}
\begin{align}\label{eqn:NLCcircEW}
&
\mcT_{\text{NLC}}(\varphi,\varphi^\prime)
e^{
-
i
z
\sin(\varphi - \vartheta)
+
i
z^\prime
\sin(\varphi^\prime - \vartheta)
}
= \sum_{n, n^\prime = -\infty}^{\infty}
e^{
-
i n \varphi
+
i n^\prime \varphi^\prime
+
i (n - n^\prime) \vartheta
}
\bigg\{
-
2
m^2
J_{n}(z)
J_{n^\prime}(z^\prime)
\\
\nonumber
&
-
m^2
\xi^2
\bigg(
1
+
\frac{s^2}{2(1 - s)}
\bigg)
\bigg[
\bigg(
f^2\Big(\frac{\varphi}{\Phi}\Big)
+
f^2\Big(\frac{\varphi^\prime}{\Phi}\Big)
\bigg)
J_{n}(z)
J_{n^\prime}(z')
-
f\Big(\frac{\varphi}{\Phi}\Big)
f\Big(\frac{\varphi^\prime}{\Phi}\Big)
\Big[
J_{n + 1}(z)
J_{n^\prime + 1}(z')
+
J_{n - 1}\big(z)
J_{n^\prime - 1}(z^\prime)
\Big]
\bigg]
\bigg\}
\;.
\end{align}
\end{widetext}
\noindent Observe that the only dependence on the angle $\vartheta$ is through
the term $\exp[i (n - n^\prime) \vartheta]$ (recall $G(\varphi)$ is also
independent of $\vartheta$, c.f. (\ref{def:G})). The integral over this angle
can then be performed, giving a $\delta$-function which means the probability
only has support on $n = n^\prime$, reducing the complexity from a doubly
infinite sum to a single one. (It is interesting to note that in the calculation
for a monochromatic plane wave~\cite{Berestetsky:1982aq} this factor setting $n
= n^\prime$ comes instead from a phase integral.)

The probability (\ref{eqn:NLC}) still has a complicated form, with two phase
integrals and an integral over the transverse momentum variable
$|\bm{\mcL}_{\perp}|^2$ which resides in the argument $z(\varphi)$ of the Bessel
functions. As the integrals cannot be done analytically, the route forward now
is to introduce a local expansion. We switch to the sum and difference variables
(\ref{def:Phase}) and expand in $\theta = \varphi - \varphi^\prime \ll 1$, once
again ignoring all derivatives of the field profile $f(\varphi/\Phi)$. Then we
have $f(\varphi/\Phi) \approx f(\varphi^\prime/\Phi) \approx f(\phi/\Phi)$, and
consequently
\begin{align}
z(\varphi) \approx z(\varphi^\prime) \approx& z(\phi)
\;,
\end{align}
(where we recall $\phi = (\vphi+\vphi')/2$). Finally, after setting $n = n^\prime$ as discussed above, the remaining terms in
the exponential are given by
\begin{align}\label{eqn:ExponentTheta}
&
G(\varphi)
-
G(\varphi^\prime)
-
n\varphi
+
n^\prime
\varphi^\prime
\nonumber\\
&=
\bigg[
\frac{s}{2 \eta_{e} (1 - s)}
\bigg(
1
+
\frac{|\bm{\mcL}_\LCperp|^2}{s^2 m^2}
+
\xi^2
f^2\Big(\frac{\phi}{\Phi}\Big)
\bigg)
-
n
\bigg]
\theta
\;.
\end{align}
The only dependence on the phase variable $\theta$ now comes from
(\ref{eqn:ExponentTheta}), which appears in the exponent of the integrand. The
integral over $\theta$ yields another $\delta$-function, so 
\begin{widetext}
\begin{align}\label{eqn:NLCalmost}
\bbP_{\trm{NLC}}^{\text{(circ)}}
=&
-
\frac{\alpha}{\eta_{e}}
\!
\int
\!
\ud\phi
\!
\int
\!
\ud s
\!
\int
\!
\ud|\bm{\mcL}_{\perp}|^2
\;
\sum_{n = -\infty}^{\infty}
\delta\left(
|\bm{\mcL}_\LCperp|^2
-
m^2
\Big[
2 \eta_{e} (1 - s)s
n
-
s^2 
\Big(
1
+
\xi^2
f^2\Big(\frac{\phi}{\Phi}\Big)
\Big)
\Big]
\right)
\nonumber\\
&
\times
\bigg\{
J_{n}^2\big(z(\phi)\big)
+
\half
\xi^2
f^2\Big(\frac{\phi}{\Phi}\Big)
\bigg(
1
+
\frac{s^2}{2(1 - s)}
\bigg)
\Big[
2
J_{n}^2\big(z(\phi)\big)
-
J_{n + 1}^2\big(z(\phi)\big)
-
J_{n - 1}^2\big(z(\phi)\big)
\Big]
\bigg\}
\;.
\end{align}
The remaining momentum integral is now trivial, giving the final result of
\begin{align}\label{eqn:NLCcirc}
\bbP_{\text{NLC}}^{\text{(circ)}}
\simeq
-
\frac{\alpha}{\eta_{e}}
\int
\ud\phi
\sum_{n = 1}^{\infty}
\int_{0}^{s_{n,*}(\phi)}
\!
\ud s
\bigg\{
J_n^2[z(\phi)]
+
\frac{1}{2}
\xi^2
f^2\Big(\frac{\phi}{\Phi}\Big)
\bigg(
1
+
\frac{1}{2}
\frac{s^2}{(1 - s)}
\bigg)
\bigg[
2
J_n^2[z(\phi)]
-
J_{n + 1}^2[z(\phi)]
-
J_{n - 1}^2[z(\phi)]
\bigg]
\bigg\} \, , 
\end{align}
\end{widetext}
in which
\begin{align}\label{eqn:zNLC}
z(\phi)
=&
\frac{
	2 n \xi \big|f\!\big(\frac{\phi}{\Phi}\big)\big|
}{
	\sqrt{
		1
		+
		\xi^2
		f^2\!\big(\frac{\phi}{\Phi}\big)
	}
}
\left[
\frac{s}{s_{n}(1 - s)}
\bigg(
1 - \frac{s}{s_{n}(1 - s)}
\bigg)
\right]^{1/2}
\end{align}
and
\begin{align}\label{def:snNLC}
s_{n}
=&
\frac{2 n \eta_{e}}{1
	+
	\xi^2
	f^2\big(\frac{\phi}{\Phi}\big)}
\;,
\quad&
\eta_{e}
=&
\frac{k \cdot p}{m^2}
\;,
\quad&
s 
=&
\frac{k \cdot \ell^\prime}{k \cdot p}
.
\end{align}
(We have suppressed the argument of $s_{n}=s_{n}(\phi/\Phi)$ for brevity.)
Momentum conservation leads to the condition $n \geq 1$ on the harmonic number,
whereas kinematic considerations lead to a phase-dependent upper bound on the $s$-integration
of $s_{n,*}(\phi) = s_n(\phi)/(1+s_n(\phi))$.  So for a circularly polarised plane wave
field, (\ref{def:GaugeCirc}), the LMA gives a direct generalisation of the
infinite monochromatic plane wave result (see e.g.  \cite{Berestetsky:1982aq}),
where the field strength has been localised, i.e.\ turned into a function of
phase, $\phi$, by replacing  $\xi \to \xi f(\phi/\Phi)$.  As mentioned in the
main text, this ad-hoc replacement has been used in the
literature~\cite{Titov:2019kdk}, but to the best of our knowledge, the necessary
approximations required, and in fact the validity of the approach, has not been
studied. In \cite{Titov:2019kdk} this trick of localising the field strength in
the monochromatic result is also used for the case of a linearly polarised plane
wave pulse. However, we will see below that the validity of this replacement may
not be applicable in all cases.

\subsection{Linearly polarised plane wave pulse}

The derivation of the LMA for a linearly polarised plane wave pulse mostly
follows the same path as for circular polarisation, and so we just point out the
key differences, most importantly the reasons why it is not possible to simply
take the standard results for the case of a linearly polarised monochromatic
plane wave (see e.g.  \cite{Ritus:1985}) and ``localise'' the field strength
$\xi$.

We will assume the pulse to be linearly polarised in the $\varepsilon$ direction
by adopting (\ref{def:Gauge}) with $\delta = 0$, hence
\begin{align}
a_\mu(\varphi)
=&
m
\xi
f\Big(\frac{\varphi}{\Phi}\Big)
\varepsilon_\mu
\cos\varphi
\;.
\end{align}
Choosing a linearly polarised background leads to additional structure in the
final expression for the probability. The source of this is quite simple: in a
circularly polarised pulse the term (\ref{eqn:GaugeSqCirc}) quadratic in the
gauge potential is slowly-varying with $\varphi$, but in the linear
case\footnote{In the monochromatic limit, $f \rightarrow 1$, the use of linear
polarisation is also well known to add some additional complexity to the
probability as the quadratic term in a circularly polarised pulse is
\emph{constant}, while it varies with the phase in the linear case. Compare
e.g.\ the results of \cite{Ritus:1985} (linear) with~\cite{Berestetsky:1982aq}
(circular) for nonlinear Compton scattering.} contains rapidly oscillating
parts,
\begin{align}
a^2(\varphi)
=
-
m^2
\xi^2
f^2\Big(\frac{\varphi}{\Phi}\Big)
\cos^2\varphi
\;.
\end{align}

The appearance of the $\cos^2 \varphi$ term means that we must implement
\emph{both} slowly-varying envelope approximations, (\ref{eqn:LinearIntApprox})
and (\ref{eqn:QuadIntApprox}), for the linear terms. The exponent in
(\ref{eqn:Reduced}) is then
\begin{align}\label{eqn:ExponentLin}
\int^{\varphi}_{-\infty}
\frac{\lp \cdot \pi_p}{k \cdot (p - \lp)}
&\simeq
G(\varphi)
-
\alpha(\varphi)
\sin\varphi
+
\beta(\varphi)
\sin 2\varphi
\;,
\end{align}
where we have introduced the function 
\begin{align}\label{def:Flin}
G(\varphi)
=&
\frac{s}{2 \eta_{e} (1 - s)}
\bigg[
\bigg(
1
+
\frac{|\bm{\ell}_\LCperp^\prime - s\bm{p}_{\LCperp}|^2}{s^2 m^2}
\bigg)
+
\frac{\xi^2}{2}
f^2\Big(\frac{\varphi}{\Phi}\Big)
\bigg]
\varphi
\;,
\end{align}
and the abbreviations
\begin{align}
\alpha(\varphi)
=&
\frac{\xi f(\frac{\varphi}{\Phi})
}{\eta_{e} m (1 - s)}
((\lp - s p) \cdot \varepsilon)
\;,
\nonumber\\
\beta(\varphi)
=&
\frac{s}{8 \eta_{e} (1 - s)}
\xi^2f^2\Big(\frac{\vphi}{\varPhi}\Big).
\end{align}
Notice that $\alpha(\cdot) \equiv \alpha_c(\cdot)$ (see (\ref{def:Alphas})) and
thus depends on the projection of the photon momentum along $\varepsilon$, but
that $\beta(\varphi)$ is \emph{independent} of the perpendicular directions.
These simple observations have far reaching consequences. Most notably, the two
trigonometric terms in (\ref{eqn:ExponentLin}) cannot be combined as was done in
(\ref{eqn:ReducedCirc}).  Therefore, each of the oscillating terms in the
exponential will have to be expanded individually, once they have been put into
the form of the Bessel generating function (\ref{def:Generating}). Furthermore,
after implementing the expansion into Bessel harmonics, the probability will
depend on terms like $J_{n}[\alpha(\varphi)]$, the argument of which depends on
both the magnitude $|\bm{\mcL}_{\LCperp}|$ and the angle $\vartheta$ (using the
notation of the circularly polarised case). As such, only \emph{one} of the two
integrals coming from the perpendicular components of the outgoing photon
momentum can be done, and the result will have a residual angular dependence (if
one chooses to do the integral in $|\bm{\mcL}_{\LCperp}|$). Remember that for
circular polarisation the simple dependence on the angle $\vartheta$ in
(\ref{eqn:NLCcircEW}) meant that the integral over $\vartheta$ could be
performed, and the probability then only had support on $n = n^\prime$. This is
not the case for linear polarisation, and one finds that the number of harmonic
sums cannot be reduced to the same amount as for the case of an infinite
monochromatic plane wave.

With all this in mind we jump ahead to the probability, expand in the phase
difference variable, $\theta = \varphi - \varphi^\prime$. and perform all the
remaining integrals which can be done analytically. Defining the combinations of
Bessel functions
\begin{widetext}
\begin{align}
\Gamma_{0,n}(\phi)
\equiv&
\sum_{r = - \infty}^{\infty}
J_{n + 2 r}\left[\alpha(\phi)\right]
J_{r}\left[\beta(\phi)\right]
\;,
\\
\Gamma_{1,n}(\phi)
\equiv&
\half
\sum_{r = - \infty}^{\infty}
\left\{
J_{n + 2r + 1}\left[\alpha(\phi)\right]
+
J_{n + 2r - 1}\left[\alpha(\phi)\right]
\right\}
J_{r}\left[\beta(\phi)\right]
\;,
\\
\Gamma_{2,n}(\phi)
\equiv&
\frac{1}{4}
\sum_{r = - \infty}^{\infty}
\left\{
J_{n + 2 r + 2}\left[\alpha(\phi)\right]
+
J_{n + 2 r - 2}\left[\alpha(\phi)\right]
+
2
J_{n + 2r}\left[\alpha(\phi)\right]
\right\}
J_{r}\left[\beta(\phi)\right]
\;,
\end{align}
with arguments
\begin{align}
\alpha(\phi)
=&
-
\frac{(n + n^\prime) \xi
	\big|f\big(\frac{\phi}{\Phi}\big)\big|\cos\vartheta}{\sqrt{1 +
	\frac{\xi^2}{2} f^2\big(\frac{\phi}{\Phi}\big)}}
\sqrt{
	w_{n + n^\prime}
	\Big(
	1 -
	w_{n + n^\prime}
	\Big)
}
,
\quad
\beta(\phi)
=
\frac{\xi^2 f^2\big(\frac{\phi}{\Phi}\big)}{8 \eta_{e}}
\frac{s}{1 - s}
\end{align}
and abbreviations
\begin{align}
w_{n + n^\prime} = \frac{s}{s_{n + n^\prime} (1 - s)} \; ,  \quad s_{n + n^\prime}
=&
\frac{(n + n^\prime) \eta_{e}}{1
	+
	\frac{\xi^2}{2}
	f^2\big(\frac{\phi}{\Phi}\big)}
\; ,
\end{align}
the probability for linearly polarised nonlinear Compton scattering finally
becomes
\begin{align}\label{eqn:NLClin}
\bbP_{\text{NLC}}^{(lin)}
&\simeq
\frac{\alpha}{2\pi \eta_{e}}
\int
\ud \phi
\sum_{n = 1}^{\infty}
\sum_{n^\prime = 1}^{\infty}
\int_0^{s_{n + n^\prime,\ast}(\phi)}
\!
\!
\ud s
\int_0^{2\pi}
\ud \vartheta
\exp\left(
-
i 
\big(n - n^\prime\big) \phi
\right)
\nonumber\\
&
\times
\bigg\{
-
\Gamma_{0,n}\big(\phi\big)
\Gamma_{0,n^\prime}\big(\phi\big)
\nonumber\\
&
-
\frac{1}{2}
\xi^2
f^2\Big(\frac{\phi}{\Phi}\Big)
\bigg(
1
+
\frac{s^2}{2(1 - s)}
\bigg)
\bigg[
\Gamma_{2,n}\big(\phi\big)
\Gamma_{0,n^\prime}\big(\phi\big)
+
\Gamma_{0,n}\big(\phi\big)
\Gamma_{2,n^\prime}\big(\phi\big)
-
2
\Gamma_{1,n}\big(\phi\big)
\Gamma_{1,n^\prime}\big(\phi\big)
\bigg]
\bigg\}
\;,
\end{align}

\noindent where the upper bound on the integration over $s$ is given by,
$s_{n+n^{\prime},*}(\phi) = s_{n+n^{\prime}}(\phi)/(1+s_{n+n^{\prime}}(\phi))$.
\end{widetext}

Due to the additional structure and the infinite summations it is not possible
to simply take the standard monochromatic plane wave result and ``localise'' the
field strength, as could be done in the circularly polarised case. In principle,
the appearance of this extra structure opens up the possibility of interference
between different harmonics. However, in the parameter regime investigated in
the main text we did not find a case where this interference was significant.

\subsection{Nonlinear Breit-Wheeler Pair Production}

Nonlinear Breit-Wheeler pair production
\cite{Breit:1934zz,Toll:1952rq,Burke:1997ew} is the decay of a photon, of
momentum $\ell$ and polarisation $\varepsilon$, into an electron-positron pair,
with momenta $\pp$ and $\qp$ respectively: $\gamma(\ell) \to e^{-}(\pp) +
e^{+}(\qp)$. In vacuum, this process is forbidden as it violates energy-momentum
conservation, but here is made possible through the interaction with a
background electromagnetic field.  Again, the amplitude is given in terms of the
Volkov wave functions (\ref{def:Volkov}), namely
\begin{align}
S_{\text{BW}}
=
-
i
e
\int
\ud^4 x
\bar{\Psi}_\pp(x)
\slashed{\epsilon}_\ell
\Psi_{-\qp}(x)
e^{- i \ell \cdot x}
.
\end{align}
The derivation of the LMA is exactly the same as for nonlinear Compton
scattering, and so we do not give any details here. Instead, we simply state the
final result for the case of the circularly polarised plane wave
(\ref{def:GaugeCirc}), namely
\begin{widetext} 
\begin{align} \label{eqn:LMABW}
        \bbP_{\text{BW}}^{(circ)}
	\simeq& 
	\frac{\alpha}{\eta_{\gamma}} 
	\int \! \ud\phi 
\sum_{n>\lceil n_{\star}(\phi) \rceil }^{\infty}
        \int_{r_{-}(\phi)}^{r_{+}(\phi)} \! \!\ud r
	\bigg\{ 
		J_{n}^2\big(z(\phi)\big) 
		- 
		\half \xi^2 f^2\Big(\frac{\phi}{\Phi}\Big) 
		\bigg( \frac{1}{2 r (1 - r)} - 1 \bigg) 
		\bigg[ 
			2 J_{n}^2\big(z(\phi)\big) 
			- 
			J_{n + 1}^2\big(z(\phi)\big)
			- 
			J_{n - 1}^2\big(z(\phi)\big) 
		\bigg] 
	\bigg\} 
\end{align}
where we have employed the abbreviations
\begin{align} \label{eqn:BWparams} 
	z(\phi) 
	=& 
	\frac{ 2 n \xi \big|f\big(\frac{\phi}{\Phi}\big)\big| }{\sqrt{1 + \xi^2
		f^2\big(\frac{\phi}{\Phi}\big)}} 
	\bigg[ 
		\frac{1}{r_n (1 - r) r} 
		\bigg( 1 - \frac{1}{r_n (1 - r) r} \bigg) 
	\bigg]^{1/2} \;, 
	\;\;
	r_{n} =& \frac{2 n\eta_{\gamma}}{1 + \xi^2 f^2} \; ,
	\;\;
	\eta_{\gamma}=& \frac{k \cdot \ell}{m^2} \; ,
	\;\;
	n_\star(\phi) =& \frac{2 (1 + \xi^2 f^2(\phi))}{\eta_{\gamma}}.
\end{align} 
\end{widetext} 
The lower bound on the harmonic number, $\lceil n_{\star}(\phi) \rceil$, is required by
    momentum conservation, as energy-momentum conservation must be satisfied when producing the
    outgoing pair. The kinematics of the nonlinear Breit-Wheeler process 
constrains the integration region to \mbox{$r_{-}(\phi) < r < r_{+}(\phi)$}, where $r_{\pm} =
\frac{1}{2}\big(1 \pm \sqrt{1 - n_{\star}(\phi)/n}\big)$.

\section{Infra-red limit ($s \to 0$) of nonlinear Compton scattering \label{app:Smalls}}

A well known discrepancy between the locally-constant field approximation and
exact QED results is the failure of the former in the ``infra-red'', $s \to 0$,
limit of the emitted photon spectrum. This is a consequence of performing a
local expansion in $\theta = \varphi - \varphi^\prime \ll 1$ for the entire
pulse, whereas the low $s$ spectrum is dominated by contributions from large
$\theta$~\cite{king15d}. Here we present a new derivation of this limit from the
full QED probability in an arbitrary plane wave pulse, and show that the same
limit can be obtained trivially in the LMA.

The probability of nonlinear Compton scattering can be expressed in differential
form as (see e.g.~\cite{Dinu:2013hsd,Ilderton:2018nws}),
\begin{widetext}
\begin{align}\label{eqn:NLCdif}
\frac{d \bbP_{\text{NLC}}}{ds}
=
-
\frac{\alpha}{\pi \eta_{e}}
\int_{-\infty}^{\infty}
\ud\phi
\int_{0}^{\infty}
\ud\theta
\;
\sin\left(
\frac{
	\theta
	\mu
}{2\eta_{e}}
\frac{s}{1 - s}
\right)
\left\{
\frac{1}{\mu}
\frac{\partial \mu}{\partial \theta}
+
\frac{g(s)}{\theta}
\big[
a(\phi + \theta/2)
-
a(\phi - \theta/2)
\big]^2
\right\}
.
\end{align}
\end{widetext}
Here we employ the Kibble effective mass $\mu$ introduced in (\ref{eqn:Kibble})
for the gauge potential  $a_\mu = (0,\bm{a})$, whence
\begin{align} \label{eqn:Kibble2}
\mu(\phi,\theta)
=
1
+
\frac{1}{\theta}
\int_{\phi - \theta/2}^{\phi + \theta/2}
\bm{a}^2
-
\bigg(
\frac{1}{\theta}
\int_{\phi - \theta/2}^{\phi + \theta/2}
\bm{a}
\bigg)^2 \; .
\end{align}
Rescaling the phase difference variable,
\begin{align}
\theta 
=
\frac{t}{s}
\; ,
\end{align}
the Kibble effective mass (\ref{eqn:Kibble2})  becomes, to leading order in
small $s$,
\begin{align}
\lim\limits_{s \rightarrow 0}
\mu
\sim
1 
+
\frac{s}{t}
\int_{-\infty}^{\infty}
\bm{a}^2
,
\end{align}
which is independent of the phase variable $\phi$. The derivative of the Kibble
mass appearing in (\ref{eqn:NLCdif}) is then trivially of order $s^2$.  Using
also that, as $s \rightarrow 0$, $g(s) \rightarrow 1/2$, and replacing
$1/(1-s)\to 1$ throughout, we find the leading order behaviour comes form the
term in small square brackets in (\ref{eqn:NLCdif}), i.e.~the squared difference
of the potentials,
\begin{widetext}
\begin{align}\label{eqn:NLCdifslim2}
\frac{d \bbP_{\text{NLC}}}{ds}
&\sim
\frac{\alpha}{\pi \eta_{e}}
\int_{-\infty}^{\infty}
\ud\phi
\int_{0}^{\infty}
\frac{\ud t}{t}
\;
\sin\left(
\frac{t}{2\eta_{e}}
\right)
\big[
\bm{a}^2(\phi + t/2s)
-
\bm{a}(\phi + t/2s)
\cdot
\bm{a}(\phi - t/2s)
\big]
,
\end{align}
\end{widetext}
in which we have also shifted integration variables to compactify the
expression.  To proceed, we Fourier transform the gauge potentials, make use of
\begin{align}
\int
\ud\phi \, 
e^{i \omega \left(\phi + t/2s \right)}
e^{i \nu \left(\phi \pm  t/2s \right)}
=
2\pi
\delta(\omega + \nu)
e^{it (\omega \mp \nu)/2s}
,
\end{align}
to get rid of the integral over $\phi$, and put the differential probability in
the form
\begin{align}\label{eqn:NLCdifslim3}
\frac{\ud \bbP_{\text{NLC}}}{\ud s}
&\sim
\frac{\alpha}{2\pi^{2} \eta_{e}}
\int
\ud\omega
\;
\bm{a}(\omega)
\cdot
\bm{a}^{\star}(\omega)
\nonumber\\
    &\times
\int_{0}^{\infty}
\frac{\ud t}{t}
\sin\left(
\frac{t}{2\eta_{e}}
\right)
\left[
1
-
\cos\left(\frac{t \omega}{s}\right)
\right]
.
\end{align}

The remaining integral over $t$ can now be performed exactly, 
\begin{align}
&
\int_{0}^{\infty}
\frac{\ud t}{t}
\sin\left(
\frac{t}{2\eta_{e}}
\right)
\left[
1
-
\cos\left(\frac{t \omega}{s}\right)
\right]
\nonumber\\
&=
-
\frac{\pi}{4}
\left[
- 
2
+
\text{sign}\left(
1 - \frac{|\omega|}{s}
\right)
+
\text{sign}\left(
1 + \frac{|\omega|}{s}
\right)
\right]
\; .
\end{align}
In the limit $s \rightarrow 0$ this yields 
\begin{align}
\lim\limits_{s \rightarrow 0}
\int_{0}^{\infty}
\frac{\ud t}{t}
\sin\left(
\frac{t}{2\eta_{e}}
\right)
\left[
1
-
\cos\left(\frac{t \omega}{s}\right)
\right]
=
\frac{\pi}{2} 
,
\end{align}
so that the infra-red limit finally becomes
\begin{align}\label{eqn:NLCdifslim4}
\lim\limits_{s \rightarrow 0}
\frac{\ud \bbP_{\text{NLC}}}{\ud s}
&=
\frac{\alpha}{4\pi \eta_{e}}
\int
\ud\omega
\;
\bm{a}(\omega)
\cdot
\bm{a}^{\star}(\omega)
\; .
\end{align}
This agrees with the result found in \cite{DiPiazza:2017raw}.

Now, the locally-constant field approximation is well known to fail in
predicting the correct value for the $s \to 0$
limit~\cite{Khok1,Khok2,king15d,Dinu:2015aci,DiPiazza:2017raw,Ilderton:2018nws,Blackburn:2018sfn}.
We have demonstrated in the main text that, on the other hand, the LMA gives a
perfect match to the exact QED results numerically. It turns out that in the
LMA, recovering the correct limit (\ref{eqn:NLCdifslim4}) is completely trivial.

First consider the LMA of nonlinear Compton scattering in a circularly polarised
plane wave pulse (\ref{eqn:NLCcirc}). The argument of the Bessel functions
$z(\phi)$, defined in (\ref{eqn:zNLC}), has leading order behaviour $z(\phi) \to
0$ in the $s \to 0$ limit. In the $z \to 0$ limit, the Bessel functions obey,
\begin{align}
\lim\limits_{z \rightarrow 0}
J_m(z)
=
\bigg\{
\begin{array}{c}
1~\text{for}~m = 0\\
0~\text{for}~m \ne 0
\end{array}
\;.
\end{align}
So the only term that remains non-zero in the $s \to 0$ limit is the term
$\propto J_{n - 1}^2(z)$ in (\ref{eqn:NLCcirc}), with $n = 1$. Then, after
Fourier transforming the remaining terms and calculating the resulting trivial
integrals, one recovers precisely (\ref{eqn:NLCdifslim4}).  The same argument
carries through for linear polarisation as well.

\section{High-field limit ($\xi \gg 1$) of the LMA \label{app:LCFA}}

We noted in the main text that the locally-constant field approximation can be
derived as the high-field limit of the LMA; we show this explicitly here.  We
will focus on the simplest case, nonlinear Compton scattering in a circularly
polarised background field, c.f.\ (\ref{eqn:NLCcirc}).

We begin by considering the behaviour of the argument of the Bessel function
(\ref{eqn:zNLC}) for $\xi \gg 1$ which should be real and positive. For a
particular value of the light-front momentum fraction, $s$, there is a minimum
value of the harmonic number given by
\begin{align}\label{eqn:nmin}
	n_{\text{min}} 
	=&
	\frac{\bar{\xi}^{3}}{2 \chi_{e}}
	\frac{s}{1 - s}
	\left[
		1 + \frac{1}{\bar{\xi}^{2}}	
	\right] 
	\;,
\end{align}
where we use the shorthand $\bar{\xi}=\xi f$ and defined $\chi_{e} = \bar{\xi}
\eta_{e}$.

We note that as $\bar{\xi}\to \infty$, $n\sim \bar{\xi}^{3}/\chi_{e}$, and hence
the corresponding harmonic order at a fixed value of $s$ becomes very large. In
this limit, the behaviour of the Bessel function terms may be determined as
follows. Let $v=s/s_{n,\ast}$ where $s_{n,*} = s_{n}/(1+s_{n})$ is the edge of
the $n$th harmonic. This removes any dependency on $n$ from the $s$
integration range: 
\[
\int_{0}^{s_{n,\ast}}\ud s \to s_{n,\ast} \int_{0}^{1} \ud v.  
\] 
Then the argument of the Bessel functions becomes:
\begin{align}\label{eqn:NLCzAlt} 
	z =& \frac{2 n \bar{\xi}}{\sqrt{1 + \bar{\xi}^{2}}} 
	\left[
		\frac{s_{n,\ast}}{s_{n}}
		\frac{v}{1-vs_{n,\ast}}
		\left(
			1-\frac{s_{n,\ast}}{s_{n}}\frac{v}{1-vs_{n,\ast}}
		\right)
	\right]^{1/2}.  
\end{align} 
Recalling that $s_{n} = 2n\eta_{e}/(1+\bar{\xi}^{2})$, we see that, in the limit of
$\bar{\xi}\to \infty$, keeping all other variables fixed, $z \to n \zeta(v)$,
where $\zeta$ is independent of $n$. In the high-field limit, $\bar{\xi} \gg 1$,
the function $\zeta(v)$ tends to 
\begin{align}\label{eqn:zetalim}
	\lim\limits_{\bar{\xi}\gg 1} 
	\zeta(v) 
	\simeq& 
	2 \left[ 
		v (1 - v)
	\right]^{1/2} \;.  
\end{align} 
In this limit, the main contribution to the probability comes from the vicinity
of $\zeta \sim 1$ or $v \sim 1/2$. In other words, the main contribution comes
from the region where $z \sim n$. Using the high field limit of $n$,  the argument of the Bessel functions, $z$, can be shown to approach the
finite value
\begin{equation} 
	z \to \frac{\bar{\xi}^{3}}{\chi_{e}} \frac{s}{1 - s} \; .
\end{equation}

To proceed we follow the approach of Ritus~\cite{Ritus:1985} and introduce a new
parameter,
\begin{align}\label{eqn:tau}
	\tau
	=&
	\frac{\bar{\xi}}{2} 
	\left[
		\left(
			\frac{z\chi_{e}}{\bar{\xi}^{3}}\frac{(1 - s)}{s} 	
		\right)^{2} 
		- 
		1
	\right]
	\;,
\end{align}
which characterises the difference between $z$ and its limiting high-field value
at the points of maximum contribution to the probability (with a normalisation
set for later convenience). Then, using the relationship between $z$ and the
harmonic number $n$, we can express $n$ in terms of $\tau$,
\begin{align}\label{eqn:ntau}
	n =& 
	\frac{\bar{\xi}^{3}}{2 \chi_{e}}\frac{s}{1 - s}
	\left(
		1 + \frac{2 \tau}{\bar{\xi}}	
	\right)
	+
	n_{\text{min}} 
	\;.
\end{align}

In the probability one now exchanges the order of summation (over $n$)  and
integration (over $s$). In the high-field limit, one can replace  the summation
by an integration over $\tau$ such that,
\begin{widetext}
\begin{align}\label{eqn:ntotau}
	\bbP_{\text{NLC}}
	\simeq&
	-
	\frac{\alpha}{\eta_{e}}
	\int \ud \phi
	\int_{0}^{\infty} \ud s
	\int_{- \bar{\xi}/2}^{\infty}
	\ud \tau
	\left(
	\frac{\bar{\xi}^{2}}{\chi_{e}} 
	\frac{s}{1 - s} 
	\right) 
	\bigg\{
		J_n^2(n \zeta)
		-
		\bar{\xi}^2
		\bigg(
			1
			+
			\frac{1}{2}
			\frac{s^{2}}{(1 - s)}
		\bigg)
		\bigg[
			\frac{1 - \zeta^{2}}{\zeta^{2}} 
			J_n^2(n \zeta)
			+
			{J_n^{\prime}}^2(n \zeta)
		\bigg]
	\bigg\}
	\;.
\end{align}
\end{widetext}
As noted previously, in the high field limit, the minimum value $n_\mathrm{min}$
for the harmonic number  becomes large, and the main contribution to the
probability comes from the regions where $\zeta \sim 1$. These two conditions
allow us to use the Watson representation~\cite{watson1995treatise} of the
Bessel functions,
\begin{align}\label{eqn:BesselApprox}
	J_{n}(n \zeta)
	\simeq&
	\Big(\frac{2}{n}\Big)^{1/3}
	\text{Ai}(y)
	\;,
	\quad&
	y
	=&
	\Big(\frac{n}{2}\Big)^{2/3}
	\Big(
		1
		-
		\zeta^2
	\Big) 
	\;,
\end{align}
where $\text{Ai}(y)$ is an Airy function. Implementing the Watson
representation, expanding around $\bar{\xi}\gg 1$ and defining $u = s/(1-s)$ we
can approximate
\begin{align}\label{eqn:yapprox}
	y\simeq&	
	\Big(
		\frac{u}{2 \chi_{e}}	
	\Big)^{2/3} 
	(1 + \tau^{2}) 
	,
\end{align}
such that the probability turns into
\begin{widetext}
\begin{align}\label{eqn:airyprob}
	\bbP_{\text{NLC}}
	\simeq&
	-
	\frac{2\alpha}{\eta_{e}}
	\int \ud \phi
	\int_{0}^{\infty} 
	\frac{\ud u}{(1 + u)^{2}}	
	\int_{0}^{\infty}
	\ud \tau
	\Big(\frac{u}{\chi_{e}}\Big)^{1/3} 
	\bigg\{
		\text{Ai}^{2}(y)  
		-
		\Big(\frac{2\chi_{e}}{u}\Big)^{2/3} 
		\bigg(
			1
			+
			\frac{1}{2}
			\frac{u^{2}}{(1 + u)}
		\bigg)
		\Big[
			y \text{Ai}^{2}(y)  
			+
			{\text{Ai}^{\prime}}^{2}(y)  
		\Big]
	\bigg\}
	\;.
\end{align}
\end{widetext}
In (\ref{eqn:airyprob}) we made use of the fact that the only dependence of the
probability on $\tau$ is through (\ref{eqn:yapprox}), and so the integration in
$\tau$ is symmetric in the $\bar{\xi}\gg 1$ limit. 

Next, we change variables to $T = (u / 2 \chi_{e})^{2/3} \tau^{2}$, use the
two Airy function identities~\cite{watson1995treatise,Ritus:1985}
\begin{align}\label{eqn:airyiden}
	y \text{Ai}^{2}(y)  
	+
	{\text{Ai}^{\prime}}^{2}(y)  
	&=
	\frac{1}{2}
	\frac{\ud^{2}}{\ud y^{2}}
	\text{Ai}^{2}(y) \; ,  \\
	\int_{0}^{\infty} 
	\frac{\ud T}{\sqrt{T}}
	\text{Ai}^{2}(A + T) 
	&=
	\frac{1}{2} 
	\int_{2^{2/3} A}^{\infty} 
	\ud x
	\text{Ai}(x) 
	\;,
\end{align}
and define 
\begin{align}\label{eqn:zbar}
	\bar{z}=& \Big(\frac{u}{\chi_{e}}\Big)^{2/3} 	
	\quad&
	\text{Ai}_{1}(\bar{z})
	=&
	\int_{\bar{z}}^{\infty}
	\ud x
	\text{Ai}(x) 
	\;,
\end{align}
to cast the probability into the form
\begin{widetext}
\begin{align}\label{eqn:close}
	\bbP_{\text{NLC}}
	\simeq&
	-
	\frac{\alpha}{\eta_{e}}
	\int \ud \phi
	\int_{0}^{\infty}
	\frac{\ud u}{(1 + u)^{2}}
	\bigg\{
		\text{Ai}_{1}(\bar{z})
		+
		\frac{2}{\bar{z}} 
		\bigg(
			1
			+
			\frac{1}{2}
			\frac{u^{2}}{(1 + u)}
		\bigg)
		\text{Ai}^{\prime}(\bar{z}) 
	\bigg\}
\end{align}
Finally, changing variables back to $s = u/(1+u)$ the probability can be put in
the form
\begin{equation}\label{eqn:lcfalim}
	\bbP_{\text{NLC}}
	\simeq
	-
	\frac{\alpha}{\eta_{e}}
	\int \ud \phi
	\int_{0}^{1}
	\ud s
	\bigg\{
		\text{Ai}_{1}(\bar{z})
		+
		\bigg(
			\frac{2}{\bar{z}} 
			+
			s \chi
			\sqrt{\bar{z}} 
		\bigg)
		\text{Ai}^{\prime}(\bar{z}) 
	\bigg\}
\end{equation}
This is \emph{exactly} the locally-constant field approximation, cf.\  (A.14)  of \cite{Ilderton:2018nws}, where $\chi_{\gamma} \equiv s \chi$. Hence, the locally-constant field approximation is nothing but the high-field limit of the more general LMA which has been the subject of the present paper.
\end{widetext}

\bibliographystyle{apsrev}
\bibliography{LMA-bib}

\end{document}